\documentclass[draftcls, onecolumn, 12pt]{IEEEtran}

\ifCLASSINFOpdf
\else
   \usepackage[dvips]{graphicx}
\fi
\usepackage{url}

\usepackage{graphicx}

\usepackage{amssymb}
\usepackage{latexsym}
\usepackage{color}
\usepackage{verbatim}
\usepackage{boldline}
\usepackage{multirow}
\usepackage{amsmath}
\usepackage{amsfonts}

\newcommand{\Z}{\ensuremath{\mathbb Z}}
\newcommand{\C}{\ensuremath{\mathbb C}}

\newcommand{\mB}{\mathcal{B}}

\newcommand{\mL}{\mathcal{L}}

\newcommand{\mO}{\mathcal{O}}

\newcommand{\mQ}{\mathcal{Q}}
\newcommand{\mR}{\mathcal{R}}
\newcommand{\mS}{\mathcal{S}}

\newcommand{\abu}{{\bf a}}
\newcommand{\bbu}{{\bf b}}

\newcommand{\hbu}{{\bf h}}

\newcommand{\nbu}{{\bf n}}
\newcommand{\sbu}{{\bf s}}
\newcommand{\pbu}{{\bf p}}
\newcommand{\qbu}{{\bf q}}

\newcommand{\ubu}{{\bf u}}
\newcommand{\vbu}{{\bf v}}

\newcommand{\xbu}{{\bf x}}
\newcommand{\ybu}{{\bf y}}
\newcommand{\wbu}{{\bf w}}

\newcommand{\Bbu}{{\bf B}}

\newcommand{\Hbu}{{\bf H}}

\newcommand{\Ibu}{{\bf I}}

\newcommand{\Qbu}{{\bf Q}}

\newcommand{\Sbu}{{\bf S}}

\newcommand{\Wbu}{{\bf W}}
\newcommand{\Xbu}{{\bf X}}
\newcommand{\Ybu}{{\bf Y}}
\newcommand{\Phibu}{{\bf \Phi}}

\newcommand{\phibu}{{\boldsymbol \phi}}

\newcommand{\qed}{\hfill \ensuremath{\Box}}

\newtheorem{df}{Definition}
\newtheorem{thr}{Theorem}
\newtheorem{lem}{Lemma}
\newtheorem{rem}{Remark}
\newtheorem{cor}{Corollary}

\newtheorem{exa}{Example}

\numberwithin{const2}{const}

\begin{document}

\title{Binary Golay Spreading Sequences and Reed-Muller Codes for Uplink Grant-Free NOMA}

\author{Nam Yul Yu,~\IEEEmembership{Senior Member,~IEEE}
\thanks{Copyright (c) 2017 IEEE. Personal use of this material is permitted.
However, permission to use this material for any other purposes
must be obtained from the IEEE by sending a request to pubs-permissions@ieee.org.}
}  

%

\maketitle

\begin{abstract}
Non-orthogonal multiple access (NOMA) is an emerging technology
for massive connectivity in machine-type communications (MTC). 
In code-domain NOMA, non-orthogonal spreading sequences are uniquely assigned
to all devices, where active ones attempt a grant-free access to a system.
In this paper, we study a set of user-specific, non-orthogonal, binary spreading sequences
for uplink grant-free NOMA.
Based on Golay complementary sequences,
each spreading sequence provides the peak-to-average power ratio (PAPR) of at most $3$ dB
for multicarrier transmission.
Exploiting the theoretical connection to Reed-Muller codes, 
we conduct a probabilistic analysis
to search for a permutation set for Golay sequences,
which presents theoretically bounded low coherence for the spreading matrix.
Simulation results confirm that 
the PAPR of transmitted multicarrier signals via the spreading sequences is 
significantly lower than those for random bipolar, Gaussian, and Zadoff-Chu (ZC) sequences. 
Also, thanks to the low coherence, 
the performance of compressed sensing (CS) based joint channel estimation (CE) and 
multiuser detection (MUD) using the spreading sequences
turns out to be superior or comparable to those for the other ones.
Unlike ZC sequences,
the binary Golay spreading sequences have only two phases regardless of the sequence length,
which can be suitable for low cost MTC devices.  
\end{abstract}


\IEEEpeerreviewmaketitle

\section{Introduction}

\IEEEPARstart{M}{assive} connectivity of wireless devices
is a key feature of machine-type communications (MTC)~\cite{Kunz:MTC}, 
which provides a platform
for the Internet of Things (IoT).
Unlike human-type communications (HTC), 
it is essential that MTC support a massive number of devices with low control overhead, low latency,
and low power consumption for delay-sensitive and energy efficient communications.

Non-orthogonal multiple access (NOMA)~\cite{Dai:noma, Dai:survey} has been of interest
for massive connectivity of devices in 5G wireless systems.
In code-domain NOMA, user-specific and
non-orthogonal spreading sequences
are assigned to all devices, where
active ones can reduce signaling overhead via uplink grant-free access. 
Allowing multiple devices to share common resources non-orthogonally with no scheduling,
uplink grant-free NOMA can be a promising technology for massive connectivity with low latency and high energy efficiency.

In uplink grant-free NOMA, 
we assume that transmitted signals of active devices are spread onto multiple subcarriers by their own spreading sequences,
and superimposed synchronously at a base station (BS) receiver.
Then, the BS needs to identify active devices with no grant procedure
and detect their own data from the non-orthogonal multiplexing.
In this paper,
we apply the principle of 
compressed sensing (CS)~\cite{Eldar:CS} for multiuser detection,
assuming that many devices are present, but only a few of them are active at a time.
Exploiting the \emph{sparse} activity,
the BS employs a CS-based detector 
for activity detection, channel estimation, and/or data detection~\cite{Abebe:iter}$-$\cite{Yu:blind}.
In CS-based detection, we consider a spreading matrix containing all spreading sequences as its columns.
Then, the \emph{coherence}~\cite{Eldar:CS}
of the spreading matrix should be as low as possible
for a CS-based detector to be able to carry out the missions successfully.
Therefore, it is essential to design spreading sequences carefully
such that the corresponding spreading matrix has theoretically bounded low coherence,
which ultimately guarantees reliable performance for
CS-based joint channel estimation (CE)
and multiuser detection (MUD) in uplink grant-free NOMA.

When the signals of active devices are spread 
onto multiple subcarriers by spreading sequences,
the high peak-to-average power ratio (PAPR) will
cause signal distortion due to the non-linearity of power amplifiers~\cite{Litsyn:peak},
which may deteriorate all potential benefits of multicarrier communications~\cite{No:papr}.
Indeed, the PAPR of transmitted signals has been an important research topic in uplink multicarrier transmission
and various reduction techniques~\cite{No:papr, Han:papr} have been proposed for mitigating it.
In this regard, spreading sequence design for uplink NOMA has another requirement that
each spreading sequence should have low PAPR for multicarrier transmission,
in addition to low coherence of the spreading matrix.
As a consequence, we believe that 
theoretically bounded low coherence and low PAPR of spreading sequences
are essential for the success of
CS-based joint CE and MUD in uplink grant-free NOMA. 

In literature,
many researchers have studied 
a variety of pilot or spreading sequences for non-orthogonal multiple access.
Based on algebraic codes, quasi-orthogonal sequences~\cite{Yang:qos} have been introduced
to increase the system capacity of code-division multiple access (CDMA).
Random sequences, where each element is taken from 
the Gaussian distribution, have been employed in~\cite{Liu:massive}$-$\cite{Jiang:noma} 
to theoretically guarantee reliable CS-based joint CE and MUD for uplink access.
In addition, pseudonoise (PN) and random QAM sequences
have been used for CE and/or MUD in uplink access~\cite{Du:block}, \cite{Apple:async}$-$\cite{Du:efficient}.
Although these sequences 
allow multiple access with low interference, 
their PAPR properties 
have not been discussed rigorously in the articles for multicarrier transmission.

To obtain sequences with low PAPR,
one can make a coding-theoretic approach.
Golay complementary sequences and sets~\cite{Golay:series}$-$\cite{Paterson:gen} 
exhibit theoretically bounded low PAPR for multicarrier transmission.
Other complementary sequences have been also studied in \cite{Liu:comp}$-$\cite{Wu:Z}
for PAPR reduction. 
Although they have desirable 
PAPR properties, 
Golay and other complementary sequences have never been investigated for CS-based joint CE and MUD in uplink grant-free NOMA.
Zadoff-Chu (ZC) sequences~\cite{Chu:ZC}, known as constant amplitude and zero autocorrelation (CAZAC) sequences,
also provide low PAPR for multicarrier transmission.
In 3GPP-LTE~\cite{3gpp:36.211}, ZC sequences have been adopted as preambles for random access.
However, the number of phases of ZC sequences gets larger as the sequence length increases,
which may require high implementation complexity in practice.
Therefore, it is worth studying 
\emph{binary} spreading sequences offering a
low implementation cost for MTC devices, as well as providing low PAPR and 
reliable CS-based joint CE and MUD for uplink grant-free NOMA.

In this paper, we study a set of binary Golay spreading sequences 
for uplink grant-free NOMA.
The non-sparse and non-orthogonal spreading sequences of length $M = 2^m$ 
can be uniquely assigned to overloaded devices for grant-free access.
Based on Golay complementary sequences, 
each spreading sequence provides the PAPR of at most $3$ dB for multicarrier transmission.
From the perspective of coding theory, 
each sequence is a coset of the first-order 
Reed-Muller (RM) codes \cite{Mac:ECC} in which
the coset representative is associated to a second-order Boolean function,
defined by a permutation pattern~\cite{Jedwab:RM, Paterson:gen}.
Exploiting the theoretical connection, 
we randomly search for a set of permutations
that ultimately presents theoretically bounded low coherence for the spreading matrix.
In this search, we conduct a probabilistic analysis to determine
the minimum number of random trials that produce
the spreading matrix with
optimum or sub-optimum coherence. 
This analysis reveals that the coherence of the spreading matrix
can be $\mO \left(\sqrt{\frac{1}{M}} \right)$
if the overloading factor is not too high.

Simulation results demonstrate that transmitted signals of active devices, spread onto multiple subcarriers
by the binary Golay spreading sequences, have the PAPR of at most $3$ dB,
which turns out to be significantly lower than those for 
random bipolar, random Gaussian, and ZC spreading sequences.
Thanks to the low coherence, 
the corresponding spreading matrix 
also exhibits reliable performance of CS-based joint
activity detection, channel estimation, and data detection
for uplink grant-free NOMA.
While the number of phases of ZC sequences increases as the sequence length, 
the binary Golay sequences
keep only two phases, regardless of the length. 
Thus, they can
offer a complexity benefit in implementation,
which can be suitable for low cost MTC devices.

This paper is organized as follows.
In Section II, we present fundamentals and backgrounds
for understanding the rest of this paper.
Section III describes the system model of uplink grant-free NOMA
and formulates the mathematical problem of CS-based joint CE and MUD.
In Section IV, we present a general framework for binary Golay spreading sequences
and investigate the coherence of the spreading matrix theoretically.
We also conduct a probabilistic analysis to determine the minimum number of random trials 
for a permutation set guaranteeing optimum or sub-optimum coherence.
Section V presents simulation results
to demonstrate the performance of CS-based joint CE and MUD in uplink grant-free NOMA
employing the binary Golay spreading sequences.
Finally, concluding remarks will be given in Section VI.

\section{Background}
Throughout this paper, we use the following notations and concepts.

\begin{itemize}
\item[$-$] $\Z_2 = \{0,1\}$ and $\Z_2 ^m$ is an $m$-dimensional vector space with each element of $\Z_2$. 
\item[$-$] In Boolean functions, quadratic, and symplectic matrices, 
`$+$' denotes the addition over $\Z_2$, or modulo-$2$ addition.
Elsewhere, it means the conventional addition.
\item[$-$] A matrix (or vector) is represented by
a bold-face upper (or lower) case letter.
\item[$-$] $(\cdot)^T$ denotes the transpose of a matrix (or vector).
\item[$-$] $\Xbu(:, t)$ denotes the $t$th column vector of a matrix $\Xbu$.
\item[$-$] $\Xbu ^{\mS}$ and $\xbu_{\mS}$ denote a sub-row matrix of $\Xbu$ and
a subvector of $\xbu$, respectively, indexed by an index set $\mS$.
\item[$-$] ${\rm rank} (\Xbu)$ is the rank of a matrix $\Xbu$, or the largest number of linearly
independent columns of $\Xbu$.
${\rm rank}_2 (\Xbu)$ is the rank of $\Xbu$ over $\Z_2$.
\item[$-$] $\Ibu$ is an identity matrix, where the dimension is determined in the context.
\item[$-$] ${\rm diag} (\hbu)$ is a diagonal matrix whose diagonal entries are from a vector $\hbu$.
\item[$-$] The inner product of vectors $\xbu$ and $\ybu$ is denoted by $\langle \xbu, \ybu \rangle$.
\item[$-$] The $l_p$-norm of $\xbu  = (x_1, \cdots, x_N)$ is denoted by
$ || \xbu ||_{p} = \left( \sum_{k=1} ^{N} |x_k|^p \right) ^{\frac{1}{p}} $ for
$1 \leq p < \infty $. 
$|| \xbu||_0 = \left| {\rm supp}(\xbu)  \right| $
is the number of nonzero elements of $\xbu$, where
${\rm supp}(\xbu) = \{i \mid x_i \neq 0, i=1, \cdots, N \}$ is 
the support of $\xbu$.
\item[$-$] A vector $\hbu \sim \mathcal{CN} (\mathbf{m}, \mathbf{\Sigma})$ is a circularly symmetric complex Gaussian random vector
with mean $\bf m$ and covariance $\mathbf{\Sigma}$.
\item[$-$] $ \mB(n, p)$ denotes the binomial distribution of $n$ independent Bernoulli trials
with each success probability $p$.
\end{itemize}

\subsection{Boolean Functions}
Let $\xbu = (x_1, \cdots, x_m)  $ be a binary vector with $m \ge 1$,
where $x_{l} \in \Z_2$ for $1 \leq l \leq m$.
A \emph{Boolean function}~\cite{Mac:ECC} is defined by a mapping
$f: \Z_2 ^m \rightarrow \Z_2$, i.e., 
\begin{equation}\label{eq:ANF}
f(\xbu) = f(x_1, \cdots, x_{m}) 
= \sum_{i=0} ^{2^m-1} c_i \prod_{l=1} ^{m} x_{l} ^{i_{l}}, \quad c_i \in \Z_2,
\end{equation}
where 
$i_{l} \in \Z_2$ is a coefficient of the binary representation of
$i = \sum_{l=1} ^{m} i_{l} 2^{l-1} $.
In~\eqref{eq:ANF}, the maximum value of $\sum_{l=1} ^{m} i_{l} $ with nonzero $c_i$
is called the \emph{degree} of $f$.
Associated to the Boolean function $f$, 
we define a binary vector $\abu = (a_0, \cdots, a_{2^m-1})^T $,
denoted by $\abu \leftrightarrow f$,
where $a_{i} =  f(i_1, \cdots, i_{m}) \in \Z_2$ for $ i = 0, \cdots, 2^m-1$.

\subsection{Reed-Muller Codes}
The $r$th order \emph{Reed-Muller (RM)} code of length $2^m$, denoted by $\mR(r, m)$,
is a set of all binary vectors $\abu \leftrightarrow f$,
where $f$ is a Boolean function of degree at most $r$, $0 \leq r \leq m$.
For $\xbu = (x_1, \cdots, x_{m})$, 
a codeword of the \emph{first-order} RM code $\mR (1,m)$ is associated to
$ f(\xbu) = \vbu \cdot \xbu ^T + e = \sum_{r=1} ^{m} v_r x_r +e $ 
for a given $\vbu = (v_1, \cdots, v_{m}) \in \Z_2 ^m$ and $e \in \Z_2$.

Each codeword of the \emph{second-order} RM code $\mR(2, m)$ is
associated to 
\begin{equation*}\label{eq:rm2}
f(\xbu) = \xbu \Qbu \xbu^T +\sum_{r=1} ^{m} v_r x_r + e, \quad v_r, e \in \Z_2,
\end{equation*}
where $\Qbu$ is an $m \times m$ binary upper triangular matrix that defines 
the second-order terms of $f$~\cite{Mac:ECC}, 
called a \emph{quadratic matrix} in this paper.
While $v_1, \cdots, v_m$ and $e$ run through all possible values for a fixed $\Qbu$, 
the quadratic Boolean function $f$ creates a \emph{coset}~\cite{Mac:ECC} of $\mR (1, m)$ in $\mR(2, m)$,
where the \emph{quadratic form} $\mQ(\xbu) =\xbu \Qbu \xbu^T $ corresponds to 
the \emph{coset representative}.
Associated with $\Qbu$, 
the \emph{symplectic} matrix is defined by $\Bbu = \Qbu + \Qbu ^T$~\cite{Mac:ECC},
which characterizes the coset.


\subsection{Golay Complementary Sequences}

Consider a Boolean function of quadratic form 
\begin{equation}\label{eq:qform}
\mQ_\pi(\xbu) =  \sum_{r=1} ^{m-1} x_{\pi(r)} x_{\pi(r+1)}, 
\end{equation}
where 
$\pi$ is a permutation in $\{1, \cdots, m \}$.
Note that there are $\frac{m!}{2}$ valid permutations~\cite{Jedwab:RM} 
for distinct quadratic forms of~\eqref{eq:qform}.
With the quadratic form, the Boolean function of
\begin{equation}\label{eq:golay}
f(\xbu) = \mQ_\pi(\xbu) + \sum_{r=1} ^{m} v_r x_r + e, \quad v_r, e \in \Z_2,
\end{equation}
gives a standard-form \emph{Golay complementary sequence}
$\abu $ of length $2^m$~\cite{Jedwab:RM}, i.e.,
$\abu \leftrightarrow f$. 
For a given $\pi$,
it is clear from \eqref{eq:golay} that
the Golay complementary sequences are a second-order 
coset of the first-order RM code $\mR (1, m)$~\cite{Paterson:gen}. 

\begin{table}[t!]
	\caption{A binary Golay complementary sequence of length $8$}
	\label{tb:golay_ex}
	\fontsize{8}{10pt}\selectfont
	\centering
	\begin{tabular}{ccc|cc|c|c}
		\hline
		\hline
		$x_3$ & $x_2$ & $x_1$ & $x_2x_1$ & $x_1x_3$ & $\mQ_{\pi}(x_1, x_2, x_3)$ & $f(x_1, x_2, x_3)$ \\
		\hline
		$0$   &  $0$  &  $0$  & $0$      & $0$      & $0$   & $0$ \\
		$0$   &  $0$  &  $1$  & $0$      & $0$     	& $0$   & $0$ \\		
		$0$   &  $1$  &  $0$  & $0$      & $0$      & $0$   & $1$ \\		
		$0$   &  $1$  &  $1$  & $1$      & $0$      & $1$   & $0$ \\	
		$1$   &  $0$  &  $0$  & $0$      & $0$      & $0$   & $1$ \\
		$1$   &  $0$  &  $1$  & $0$      & $1$      & $1$   & $0$ \\		
		$1$   &  $1$  &  $0$  & $0$      & $0$      & $0$   & $0$ \\		
		$1$   &  $1$  &  $1$  & $1$      & $1$      & $0$   & $0$ \\		
		\hline
		\hline
	\end{tabular}
\end{table}

\begin{exa}
For $m = 3$, let $\pi = \{2, 1, 3\}$ be a permutation in $\{1, 2, 3\}$.
Then, the quadratic form of~\eqref{eq:qform} is 
$\mQ_\pi (x_1, x_2, x_3)=x_2 x_1 + x_1 x_3$.
If $(v_1, v_2, v_3) = (0, 1, 1)$ and $e = 0$,
the Boolean function of~\eqref{eq:golay} is
$f(x_1, x_2, x_3) = x_2 x_1 + x_1 x_3 + x_2 + x_3$, 
which yields 
a binary Golay complementary sequence of length $8$, or $\abu = (0,0,1,0,1,0,0,0)^T$. 

Table~\ref{tb:golay_ex} describes this example.
\end{exa}

\subsection{Peak-to-Average Power Ratio (PAPR)}
Let $\abu = (a_0, \cdots, a_{M-1})^T$ be a binary sequence of length $M$, where $a_i \in \Z_2$.
Then, its \emph{modulated} sequence is given by
$\bbu = \psi(\abu) = (b_0, \cdots, b_{M-1})^T$, where $b_i = (-1)^{a_i}$. 
When the modulated sequence $\bbu  $ is transmitted through $M$ subcarriers,
the peak-to-average power ratio (PAPR) of its OFDM signal is defined by~\cite{Litsyn:peak}
\begin{equation}\label{eq:papr}
{\rm PAPR} (\bbu)= \max_{t \in [0, 1) } 
\frac{ \left| \sum_{i=0} ^{M-1} b_i e^{j 2 \pi i t} \right|^2 }{ M} .
\end{equation}
In~\eqref{eq:papr},
the computation of peak power containing a continuous-time signal
can be approximated using the discrete Fourier transform (DFT) of the
oversampled, discrete signal~\cite{Litsyn:peak,Litsyn:maxima}.
It is well known that the modulated Golay complementary sequence has the PAPR of at most $2$~\cite{Jedwab:RM}.


\subsection{Compressed Sensing (CS)}
Compressed sensing (CS)~\cite{Eldar:CS} is to recover a sparse signal
from the far fewer measurements than the signal dimension.
A signal $\xbu \in \C^N$ is called \emph{$K$-sparse} if it has at most $K$ nonzero elements, 
where $K \ll N$.
In CS, the sparse signal $\xbu$ is measured by 
\begin{equation}\label{eq:smv}
\ybu = \Phibu \xbu + \nbu,
\end{equation}
where $\Phibu$ is an $M \times N$ measurement matrix with $M < N$
and $\nbu$ is the measurement noise.
If $\Phibu$ obeys the \emph{restricted isometry property (RIP)}~\cite{Candes:rip},
stable and robust reconstruction of $\xbu$
is guaranteed from $\ybu$.
The CS reconstruction is accomplished by solving an $l_1$-minimization
problem with convex optimization or greedy algorithms.

If $\Phibu$ is a random Gaussian or Bernoulli matrix, 
the RIP analysis 
presents a theoretical guarantee for CS recovery~\cite{Eldar:CS}.
However, 
the RIP analysis of $\Phibu$ is NP-hard~\cite{Pfetsch:NP} 
if it is a deterministic matrix. 
In this case, we can present a theoretical recovery guarantee
using the \emph{coherence}, defined by
\begin{equation}\label{eq:coh_org}
\mu(\Phibu) = \max_{ 1 \leq j_1 \neq j_2 \leq N  }
\frac{\left| \left \langle \phibu_{j_1} ,   \phibu_{j_2}  \right \rangle \right| }
{|| \phibu_{j_1} ||_2 || \phibu_{j_2} ||_2},
\end{equation}
where $\phibu_j$ is the $j$th column vector of $\Phibu$.
From~\cite{Eldar:CS}, we have
\begin{equation}\label{eq:spark}
{\rm spark}(\Phibu) >  1 + \mu^{-1} \left(\Phibu \right) ,
\end{equation}
where ${\rm spark}(\Phibu)$ is the smallest number of columns of $\Phibu$
that are linearly dependent.
Since unique reconstruction of a $K$-sparse signal is guaranteed 
if and only if $K < \frac{{\rm spark}(\Phibu)}{2}$~\cite{Eldar:CS},
\eqref{eq:spark} implies that the coherence of $\Phibu$ should be as low as possible 
for unique CS recovery with larger $K$. 

If multiple signals of interest are acquired by $\Phibu$,
the single measurement vector (SMV) problem of~\eqref{eq:smv} can be extended to
a multiple measurement vector (MMV) problem
of $\Ybu = \Phibu \Xbu + \Wbu$, 
where $\Xbu \in \C^{N \times J}$ is a collection of $J$ signals of interest
and $\Wbu$ is the measurement noise.
In some applications, $\Xbu$ can be \emph{jointly} $K$-sparse,
which means that the columns of $\Xbu$ 
have a common support $\mS$ with $|\mS| = K$.
Then, a necessary and sufficient condition~\cite{Chen:MMV, Davies:rank} for unique recovery of
jointly $K$-sparse signal $\Xbu$
from $\Ybu = \Phibu \Xbu$ is 
\[
K < \frac{{\rm spark} \left(\Phibu \right) - 1 + \rm{rank} \left(\Xbu \right)}{2}.
\]
Again, the lower coherence of $\Phibu$ yielding the higher ${\rm spark} \left(\Phibu \right)$
ultimately guarantees unique reconstruction of a jointly $K$-sparse signal with larger $K$.

\section{System Model}
\begin{figure}
	\centering
	\includegraphics[width=0.65\textwidth, angle=0]{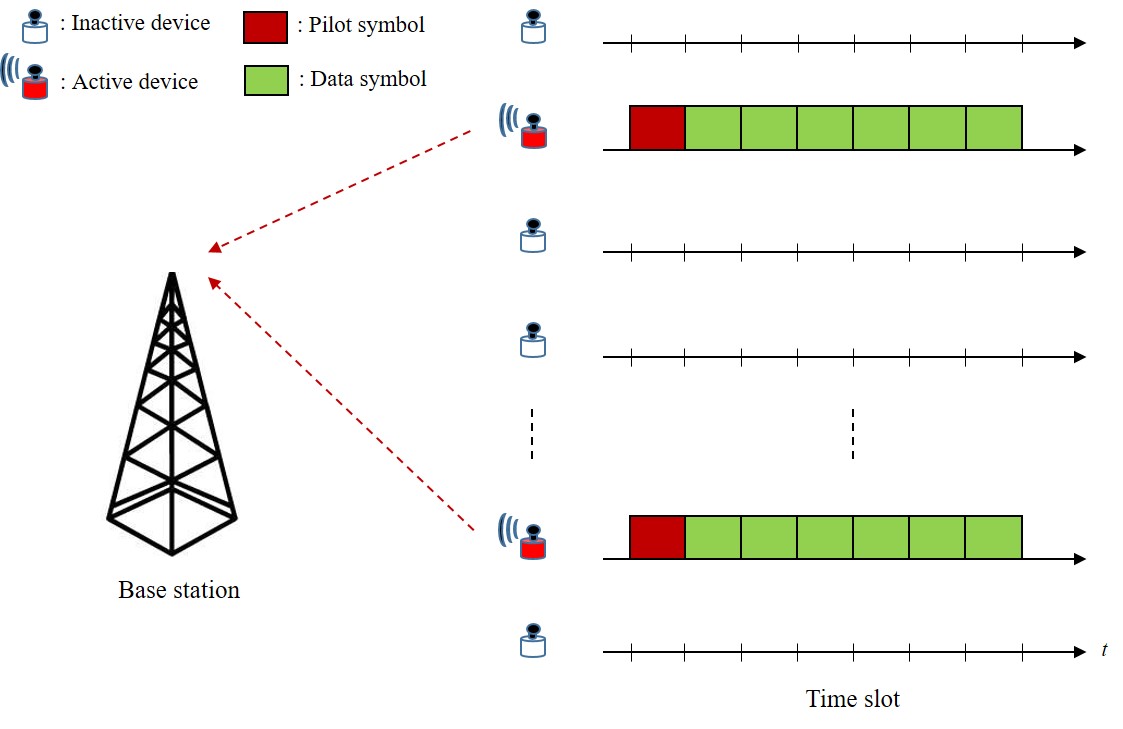}
	\caption{System model for uplink grant-free NOMA.}
	\label{fig:system}
\end{figure}

In uplink grant-free NOMA, 
we assume that a base station (BS) equipped with a single antenna
accommodates total $N$ devices each of which transmits with a single antenna.
We also assume that devices are synchronized in a frame structure
of $J$ time slots, where the activity of each device remains unchanged during an entire frame.
For massive connectivity, $N$ is large, but
the number of active devices in a frame is assumed to be far less than $N$.
Each active device sends a pilot symbol for channel estimation
in the first slot of a frame, and 
subsequently sends data symbols for the next $J-1$ slots~\cite{Du:joint}.
Figure~\ref{fig:system} illustrates the system model, where $J=7$.

At a time slot $t$ of a frame,
the transmitted symbol $u_n ^{(t)}$ of active device $n$
is spread onto $M$ subcarriers using a unique spreading sequence $\sbu_n = (s_{0,n}, \cdots, s_{M-1,n})^T $,
where 
$M<N$ in NOMA.
We assume that the spread symbols are transmitted through a flat fading channel, 
which is static, keeping the channel gain 
unchanged during an entire frame.
Then, the received signal of BS is represented 
by
\begin{equation}\label{eq:y}
\ybu^{(t)} = \sum_{n=1} ^N h_n  \sbu_n u_n ^{(t)} + \wbu ^{(t)}
=  \Sbu {\rm diag} \left(\hbu \right) \ubu ^{(t)} + \wbu ^{(t)},
\end{equation}
where $ 1 \leq t \leq J$. In~\eqref{eq:y},
$\Sbu = [\sbu_1, \cdots, \sbu_N] \in \C^{M \times N} $ is the spreading matrix,
$\hbu = (h_1, \cdots, h_N)^T \sim \mathcal{CN}(\bf{0}, \Ibu) $, 
where $h_n$ is a channel gain between device $n$ and BS, and
$\wbu^{(t)} \sim \mathcal{CN}(\textbf{0}, \sigma_n ^2 \Ibu) $ is the complex Gaussian noise vector. 
Also, $\ubu^{(t)} = (u_1 ^{(t)}, \cdots, u_N ^{(t)} )^T \in \C^N$
is a modulated symbol vector from all $N$ devices at a time slot $t$,
where $u_{n} ^{(t)} = 0$ if device $n$ is inactive. 
At $t=1$, active device $n$ transmits a pilot symbol, 
which is assumed to be $u_n ^{(1)} = 1$ for simplicity.
Finally, if $K $ ($\ll N$) devices are active, 
$\ubu^{(t)} $ is $K$-sparse.

In Figure~\ref{fig:system}, the system model
assumes that a set of active devices remains unchanged over a frame of $J$ time slots.
Under this assumption, we achieve the \emph{frame-wise joint sparsity}, formulated by
\begin{equation*}\label{eq:supp}
{\rm supp} (\ubu^{(1)})  = {\rm supp} (\ubu^{(2)} )
= \cdots = {\rm supp} (\ubu^{(J)} ) \triangleq \mS,
\end{equation*}
where 
$\mS$ is a set of active devices with
$| \mS | = K$.
Under a static channel over a frame,  
we have a multiple measurement vector (MMV) 
model of
$\Ybu = \Sbu \Xbu + \Wbu$,
where $\Ybu = [\ybu^{(1)}, \cdots, \ybu^{(J)} ]$,
$\Xbu = {\rm diag}(\hbu) \cdot [\ubu^{(1)}, \cdots, \ubu^{(J)} ]$,
and $\Wbu = [\wbu^{(1)}, \cdots, \wbu^{(J)} ]$, respectively.
Then, one can apply
a joint sparse recovery algorithm 
to obtain an estimate of $\Xbu^\mS$, or $\widehat{\Xbu} ^{\widehat{\mS}}$,
by solving the MMV problem, 
where $\widehat{\mS}$ is an estimate of $\mS$. 
Since $u_n ^{(1)} = 1$ for active device $n$, 
the channel gain vector of active devices can be estimated by 
\begin{equation}\label{eq:hhat}
\widehat{\hbu}_{\widehat{\mS}} = \widehat{\Xbu} ^{\widehat{\mS}} \left( :, 1\right),
\end{equation}
and $ \widehat{\hbu}_{\widehat{\mS} ^c} = {\bf 0}$
for $\widehat{\mS} ^c = \{1, \cdots, N\} \setminus \widehat{\mS}$. 
With $\widehat{\hbu}_{\widehat{\mS}} $, 
data symbols of active devices can be detected by 
\begin{equation}\label{eq:uhat}
\widehat{\ubu} ^{(t)} _{\widehat{\mS}} = \left({\rm diag} (\widehat{\hbu}_{\widehat{\mS}}) \right) ^{-1} \cdot \widehat{\Xbu} ^{\widehat{\mS}} \left( :, t \right),
\quad t = 2, \cdots, J, 
\end{equation}
and $\widehat{\ubu} ^{(t)} _{\widehat{\mS} ^c} = {\bf 0}$.
Finally, \eqref{eq:hhat} and \eqref{eq:uhat} complete CS-based joint channel estimation (CE) and 
multiuser detection (MUD) for uplink grant-free NOMA.

\section{Binary Golay Spreading Sequences}

\subsection{Framework}

Let $\xbu = (x_1, \cdots, x_{m})$ for a positive integer $m\ge 1$.
Define a linear Boolean function by
\begin{equation*}\label{eq:boolL}
 \mL_c(\xbu) =  \sum_{r=1} ^{m} v_r x_r , \quad v_r \in \Z_2,
\end{equation*}
where $c = \sum_{r=1} ^{m} v_r 2^{r-1} $.
Clearly, $c$ runs through $0$ to $2^m-1$
while $\vbu = (v_1, \cdots, v_{m})$ takes all possible vectors of $\Z_2 ^m$. 
Then, $\mL_c$ generates a subcode $\mO_m \subset \mR (1, m)$ while $0 \leq c \leq 2^m-1$,
where all the codewords are mutually orthogonal~\cite{Mac:ECC}.
%

With the quadratic form $\mQ_{\pi} (\xbu)$ of~\eqref{eq:qform},
a Boolean function 
$f_{\pi} ^{(c)} (\xbu) = \mQ_{\pi}(\xbu) + \mL_c (\xbu)$ is associated to
a binary Golay complementary sequence 
of length $2^m$ for any $\pi$ and $c$.
While $c$ runs through $0$ to $2^m-1$ for a given $\pi$,
$f_\pi ^{(c)} (\xbu) $ generates 
a coset of $\mO_m$ in $\mR (2, m)$ with
the coset representative $\mQ_{\pi} $. 
When each codeword of the coset is arranged as a column, 
it creates a $2^m \times 2^m$ matrix,
where a pair of distinct columns are mutually orthogonal.
Equivalently, we obtain $2^m$ orthogonal spreading sequences each of which has the length $2^m$.

To obtain non-orthogonal spreading sequences,
we expand the matrix by employing more permutations.
In what follows, we present a framework of non-orthogonal, binary Golay spreading sequences
for uplink grant-free NOMA.

\begin{df}\label{def:spread}
Let $M$ and $N$ be the length and the number of spreading sequences, respectively,
where $M = 2^m$ for $m \ge 1$.
Consider $L$ distinct permutations $\pi_1, \cdots, \pi_L$ in $\{1, \cdots, m \}$, 
where $L = \lceil \frac{N}{M} \rceil$.
Let $f_{\pi_k} ^{(c)} (\xbu) = \mQ_{\pi_k}(\xbu) + \mL_{c} (\xbu)$,
where $\abu_{\pi_k} ^{(c)} \leftrightarrow f_{\pi_k} ^{(c)} $
and $\bbu_{\pi_k} ^{(c)} = \psi(\abu_{\pi_k} ^{(c)})$ 
for $0 \leq c \leq 2^m-1$ and $1 \leq k \leq L$.
For a given $\pi_k$, we construct an $M \times M$ orthogonal matrix of
\begin{equation}\label{eq:Phibu_k1}
\Phibu_k  = \left[\bbu_{\pi_k} ^{(0)}, \  \bbu_{\pi_k} ^{(1)}, \ \cdots,  \ \bbu_{\pi_k} ^{(M-1)} \right].
\end{equation}
With $L$ permutations, we then generate an $M \times LM$ spreading matrix of
\begin{equation}\label{eq:Phibu}
\Phibu = \frac{1}{\sqrt{M}} [\Phibu_1,  \cdots,  \Phibu_L ] .
\end{equation}
Finally, choosing the first $N$ columns from $\Phibu$,
we construct the spreading matrix $\Sbu$ of~\eqref{eq:y},
where each column can be a spreading sequence of length $M=2^m$.
Since each spreading sequence 
is a binary Golay complementary sequence,
it is obvious that its PAPR is
at most $2$, or equivalently $3$ dB.
\end{df}

\begin{rem}
In 
Definition~\ref{def:spread}, it is clear that
the modulated sequences associated to $\mL_c (\xbu)$ form the Walsh-Hadamard matrix $\Hbu$ of $\pm 1$ elements
while $c$ runs through $0 $ to $2^m-1$. 
Then, if $\qbu_k \leftrightarrow \mQ_{\pi_k} (\xbu)$ and $\pbu_k = \psi(\qbu_k)$, 
it is readily checked that the submatrix $\Phibu_k$ of \eqref{eq:Phibu_k1} is equivalently generated by 
\begin{equation}\label{eq:Phik_H}
\Phibu_k =  {\rm diag} (\pbu_k) \cdot \Hbu.
\end{equation}
Thus,
the spreading matrix $\Phibu$ can be easily generated by the matrix operations of \eqref{eq:Phibu} and \eqref{eq:Phik_H}.
\end{rem}

In Definition~\ref{def:spread}, 
$\abu_{\pi_k} ^{(c)}$ is a binary spreading sequence of each element $0$ or $1$,
whereas $\bbu_{\pi_k} ^{(c)}$ is its modulated, $\pm 1$-sequence.
Throughout this paper,
we equivalently use $\abu_{\pi_k} ^{(c)} $ and $\bbu_{\pi_k} ^{(c)}$
to denote a spreading sequence.
In next subsection, we investigate the coherence of $\Phibu$,
which depends on a selection of $L$ permutations $\pi_1, \cdots, \pi_L$.

\subsection{Coherence Analysis}

As discussed in Section II.E, 
a collection of all spreading sequences, or spreading matrix must have low coherence
to guarantee reliable performance for CS-based joint CE and MUD
in uplink grant-free NOMA. 
For simplicity, we assume that $N$ is a multiple of $M$, or
$\Sbu = \Phibu$ in this paper, which leads to
$\mu \left(\Sbu \right) = \mu \left(\Phibu \right)$. 
From~\eqref{eq:coh_org},
the coherence of $\Phibu$ in Definition~\ref{def:spread}
is given by 
\begin{equation}\label{eq:coh_def}
\mu(\Phibu) = \max_{ \substack{1 \leq k_1 , k_2 \leq L \\ 0 \leq c_1, c_2 \leq M-1} }
\frac{\left| \left \langle \bbu_{\pi_{k_1}} ^{(c_1)},   \bbu_{\pi_{k_2}} ^{(c_2)} \right \rangle \right| }{M} ,
\end{equation}
where $c_1 \neq c_2$ if $k_1 = k_2$.


In what follows, we show that the coherence of $\Phibu$ 
is determined by the rank information of a symplectic matrix
corresponding to a pair of permutations.

\begin{lem}\label{lm:rm_coh}
In Definition~\ref{def:spread}, let $\pi_{k_1}$ and $\pi_{k_2}$
be a pair of permutations 
with $k_1  \neq k_2$, and $\Phibu_{k_1, k_2} =\frac{1}{\sqrt{M}} [\Phibu_{k_1}, \Phibu_{k_2}]$.
Let
$\Qbu_{k_1} $ and $ \Qbu_{k_2}$
be the quadratic matrices
corresponding to their quadratic forms $\mQ_{\pi_{k_1}}$ and $\mQ_{\pi_{k_2}}$, respectively.
Define
a symplectic matrix $\Bbu_{k_1, k_2} = \Qbu_{k_1, k_2} + \Qbu_{k_1, k_2}^T$,
where $\Qbu_{k_1, k_2} = \Qbu_{k_1} + \Qbu_{k_2}$.
Then, if ${\rm rank}_2  \left( \Bbu_{k_1, k_2} \right) = r$,
\begin{equation}\label{eq:coh_Pk}
\mu(\Phibu_{k_1, k_2}) = \frac{1}{\sqrt{2^{r}}}.
\end{equation}
\end{lem}

\noindent \textit{Proof:} 
From \eqref{eq:coh_def},
\begin{equation}\label{eq:coh_def2}
\mu(\Phibu_{k_1, k_2}) = \max_{  0 \leq c_1, c_2 \leq M-1} 
\frac{\left| \left \langle \bbu_{\pi_{j_1}} ^{(c_1)},   \bbu_{\pi_{j_2}} ^{(c_2)} \right \rangle \right| }{M} ,
\end{equation}
where $(j_1, j_2)= (k_1, k_1), (k_1, k_2)$, or $ (k_2, k_2)$, and
$c_1 \neq c_2$ if $j_1 = j_2$.
In~\eqref{eq:coh_def2}, if $j_1 = j_2$,
$ \left \langle\bbu_{\pi_{j_1}} ^{(c_1)}, \bbu_{\pi_{j_2}} ^{(c_2)} \right \rangle = 0$ with $c_1 \neq c_2$, 
as they are the column vectors of the orthogonal matrix $\Phibu_{j_1}$.

Meanwhile, if $(j_1, j_2) = (k_1, k_2)$, 
then $\abu_{\pi_{k_1}} ^{(c_1)} $ (or $\abu_{\pi_{k_2}} ^{(c_2)} $) is 
a codeword of a coset of $\mO_m$ in $\mR(2, m)$ with the coset representative $\mQ_{\pi_{k_1}}$ (or $\mQ_{\pi_{k_2}}$).
Also, $\abu_{\pi_{k_1}} ^{(c_1)} + \abu_{\pi_{k_2}} ^{(c_2)} $ 
is a codeword of a coset of $\mO_m$ in $\mR(2, m)$,
which is characterized by the symplectic matrix $\Bbu_{k_1, k_2}$.
Since the Hamming weight $w$ of $\abu_{\pi_{k_1}} ^{(c_1)} + \abu_{\pi_{k_2}} ^{(c_2)} $ 
corresponds to the inner product $2^m-2w$ of $\bbu_{\pi_{k_1}} ^{(c_1)}$ and $\bbu_{\pi_{k_2}} ^{(c_2)} $, 
the result of Theorem 5 of Ch. 15~\cite{Mac:ECC} can be translated in terms of inner product.
In other words, the theorem says 
$ \left \langle\bbu_{\pi_{k_1}} ^{(c_1)}, \bbu_{\pi_{k_2}} ^{(c_2)} \right \rangle = 0$ or $\pm 2^{m-h}$ 
if ${\rm rank}_2 \left(\Bbu_{k_1, k_2} \right) = 2h$.
Thus, if $r= {\rm rank}_2 \left(\Bbu_{k_1, k_2} \right)$,
\eqref{eq:coh_Pk} is obvious from the rank information and 
\eqref{eq:coh_def2}.
\qed


\begin{thr}\label{th:rank_coh}
From Lemma~\ref{lm:rm_coh}, 
recall $\Phibu_{k_1, k_2} =\frac{1}{\sqrt{M}} [\Phibu_{k_1}, \Phibu_{k_2}]$
and the symplectic matrix $\Bbu_{k_1, k_2}$
corresponding to the permutation pair $\pi_{k_1}$ and $\pi_{k_2}$, respectively, where $k_1 \neq k_2$.
Then, if 
\begin{equation}\label{eq:r_min}
r_{\min} = \min_{1 \leq k_1 \neq k_2 \leq L} {\rm rank}_2 \left(\Bbu_{k_1, k_2}\right),
\end{equation}
the coherence of $\Phibu$ in Definition~\ref{def:spread} is given by
\begin{equation}\label{eq:rm_coh}
\mu(\Phibu) = \frac{1}{\sqrt{2^{r_{\min}}}}.
\end{equation}
\end{thr}

\noindent \textit{Proof:} 
From Lemma~\ref{lm:rm_coh}, it is readily checked that
\begin{equation}\label{eq:coh2}
\mu(\Phibu) = \max_{ 1 \leq k_1 \neq  k_2 \leq L  }
\mu \left( \Phibu_{k_1, k_2} \right) 
 =  \max_{ 1 \leq k_1 \neq  k_2 \leq L  } \frac{1}{\sqrt{2^{r}}},
\end{equation}
where $r = {\rm rank}_2  \left( \Bbu_{k_1, k_2} \right) $.
Obviously, \eqref{eq:rm_coh} is obtained from \eqref{eq:coh2} with $r_{\min}$ of \eqref{eq:r_min}.
\qed

From Theorem~\ref{th:rank_coh},
the following corollary is straightforward.

\begin{cor}\label{co:min_coh}
In Definition~\ref{def:spread}, the coherence of $\Phibu$ satisfies
\begin{equation*}\label{eq:min_coh}
\mu \left( \Phibu \right) \geq \left \{ \begin{array}{ll} \sqrt{\frac{2}{M}}, & \mbox{ if } m \mbox{ is odd}, \\
\sqrt{\frac{1}{M}}, &  \mbox{ if } m \mbox{ is even}, \end{array} \right.
\end{equation*}		
where $M=2^m$.
\end{cor}

\noindent \textit{Proof}: From~\cite{Mac:ECC}, Theorem 2 of Ch. 15 implies that
the rank of an $m \times m$ symplectic matrix is an even number between $0$ and $m$.
Thus, if $m$ is odd, $r_{\min}$ of Theorem~\ref{th:rank_coh} is
at most $m-1$, which yields $\mu \left(\Phibu \right) \ge \frac{1}{\sqrt{2^{m-1}}} =  \sqrt{\frac{2}{M}}$.
Similarly, if $m$ is even, $r_{\min}$ 
is at most $m$,  
where $\mu \left(\Phibu \right) \ge \frac{1}{\sqrt{2^{m}}} =  \sqrt{\frac{1}{M}}$.
\qed

Corollary~\ref{co:min_coh} shows that
the \emph{optimum} coherence of $\Phibu$ 
is $\mu_{\rm opt} ^{(\rm odd)} \left( \Phibu \right) = \sqrt{\frac{2}{M}}$ for odd $m$, 
and $\mu_{\rm opt} ^{(\rm even)} \left( \Phibu \right) = \sqrt{\frac{1}{M}}$ for even $m$, respectively.
From the proofs of Theorem~\ref{th:rank_coh} and Corollary~\ref{co:min_coh}, 
it is obvious that the sub-optimum coherence is twice of the optimum.

To obtain the spreading matrix $\Phibu$ with low coherence, 
Theorem~\ref{th:rank_coh} suggests that
one has to find a set of $L$ permutations in
which the symplectic matrix $\Bbu_{k_1, k_2}$ corresponding to
any permutation pair $\pi_{k_1}$ and $\pi_{k_2}$ has 
the largest possible rank. In next subsection, we discuss how to find such permutations.



\subsection{Permutation Search}

Let $\Gamma = \{ \pi_1, \cdots, \pi_L \}$ be a set of $L$ permutations
in Definition~\ref{def:spread}, 
where $\pi_k$ is a permutation for the submatrix $\Phibu_k$, $1 \leq k \leq L$. 
For any permutation pair $\pi_{k_1}$ and $\pi_{k_2}$ in $\Gamma$, 
where $1 \leq k_1 \neq k_2 \leq L$,
each one of  
$\abu_{\pi_{k_1}} ^{(c_1)}$ and $\abu_{\pi_{k_2}} ^{(c_2)}$
should be a codeword of the \emph{Kerdock} code~\cite{Mac:ECC},
to present a maximum possible rank for 
$\Bbu_{k_1, k_2}$. 
Equivalently, the quadratic forms $\mQ_{\pi_{k_1}}$ and $\mQ_{\pi_{k_2}}$ 
should belong to those of a Kerdock code for any permutation pair in $\Gamma$. 
In~\cite{Jedwab:RM}, it has been reported that 
the quadratic forms corresponding to some permutations may be those of a Kerdock code,
but others are not.
Similarly, Tables III-V of~\cite{Yang:qos} showed that some quadratic forms of a Kerdock code
do not correspond to those of permutations.
Therefore,
it is important to identify a permutation set $\Gamma$ in which the quadratic forms
of any permutation pair belong to those of a Kerdock code,
which results in the optimum coherence of $\Phibu$. 
To the best of our knowledge, however, 
there is no constructive way to find such permutations.
As an alternative, 
we can search for $L$ permutations exhaustively, 
but it requires $\binom{m!/2}{L}$ trials\footnote{We can reduce the search space by exploiting the fact that 
the first row of $\Bbu_{k_1, k_2}$ should be nonzero~\cite{Pott:diff}.
However, it also requires at least $\left( 2 (m-3)! \right)^L $ trials,
which can be enormous for large $m$ and $L$.}, 
which is computationally infeasible for large $m$ and $L$.

\begin{table*}[t!]
	\caption{Experimental Probability of Coherence of $\Phibu_{k_1, k_2}$}
	\label{tb:coh_pair}
	\fontsize{8}{10pt}\selectfont
	\centering
	\begin{tabular}{c|c|c|c|c||c|c|c|c|c}
		\hline
		$m$ & $M$ & $r$ & $\mu(\Phibu_{k_1, k_2})$ & $p_r$     & $m$ & $M$ & $r$ &  $\mu(\Phibu_{k_1, k_2})$ & $p_r$   \\ 
		\hline
		$5$	 & $32$  & $2$ &   $0.5$    &  $2.543926\times 10^{-1}$      		& $6$  & $64$  & $2$ &  $0.5$    &  $5.856820\times 10^{-2}$  \\
			 &       & $4$ &   $0.25$   &  $7.456074\times 10^{-1}$      		&      &       & $4$ &  $0.25$   &  $5.848499\times 10^{-1}$  \\
			 &       &     &            &                  						&      	&      & $6$ &  $0.125$  &  $3.565819\times 10^{-1}$  \\		
		\hline
		$7$  & $128$       & $2$ &  $0.5$    &  $1.310750\times 10^{-2}$       	& $8$  & $256$ & $2$ &  $0.5$    &  $2.218900\times 10^{-3}$  \\
			 &       	   & $4$ &  $0.25$   &  $2.214437\times 10^{-1}$      	&      &       & $4$ &  $0.25$   &  $5.931190\times 10^{-2}$  \\
			 &     		   & $6$ &  $0.125$  &  $7.654488\times 10^{-1}$      	&	   &       & $6$ &  $0.125$  &  $5.814335\times 10^{-1}$  \\	
			 &       &     &           &                   						&      &       & $8$ &  $0.0625$ &  $3.570357\times 10^{-1}$  \\
		\hline
		$9$  & $512$ & $2$ &  $0.5$    &  $3.352000\times 10^{-4}$       		& $10$ & $1024$ & $2$ &  $0.5$    &  $4.190000\times 10^{-5}$  \\ 
		     &       & $4$ &  $0.25$   &  $1.153890\times 10^{-2}$       		& 	   &        & $4$ &  $0.25$   &  $1.914300\times 10^{-3}$  \\
			 &       & $6$ &  $0.125$  &  $2.285468\times 10^{-1}$       		& 	   &        & $6$ &  $0.125$  &  $5.557250\times 10^{-2}$  \\		
			 &       & $8$ &  $0.0625$ &  $7.595791\times 10^{-1}$       		& 	   &        & $8$ &  $0.0625$ &  $5.863399\times 10^{-1}$  \\			
			 &       &     &           & 	     								&      &        & $10$ & $0.03125$ &  $3.561314\times 10^{-1}$  \\				 			  					 
		\hline								 		 		
	\end{tabular}
\end{table*}

In this paper, we make a probabilistic approach to
find a set of $L$ permutations that presents the optimum or sub-optimum coherence for $\Phibu$.
We begin with an experimental result for the probability that a symplectic matrix is of a particular rank.
In this experiment, 
we choose a pair of permutations $\pi_{k_1}$ and $\pi_{k_2}$ randomly from all possible ones, and then
compute the rank of the symplectic matrix $\Bbu_{k_1, k_2}$ 
corresponding to the quadratic forms $\mQ_{\pi_{k_1}}$ and $\mQ_{\pi_{k_2}}$.
Repeating with different $\pi_{k_1}$ and $\pi_{k_2}$,
we can empirically compute the probability $p_r$ of 
${\rm rank}_2  \left(\Bbu_{k_1, k_2} \right) = r$.
In terms of coherence, we can say that 
$\mu \left( \Phibu_{k_1, k_2}\right) = \frac{1}{\sqrt{2^r}}$ with probability $p_r$,
where $\Phibu_{k_1, k_2} =\frac{1}{\sqrt{M}} [\Phibu_{k_1}, \Phibu_{k_2}]$.

Table~\ref{tb:coh_pair} shows the experimental probability $p_r$ 
of $\mu \left(\Phibu_{k_1, k_2} \right) = \frac{1}{\sqrt{2^r}}$ in accordance with
${\rm rank}_2 \left( \Bbu_{k_1, k_2} \right)=r$ for $5 \leq m \leq 10$,
where the permutation pair $\pi_{k_1}$ and $\pi_{k_2}$ are chosen randomly at each
of total $10^7$ random trials.
Note that $r $ is an even integer for $2 \leq r \leq m$,
which cannot be $0$ due to $k_1 \neq k_2$.
Table~\ref{tb:coh_pair} demonstrates that
if $m$ is odd, one can choose a permutation pair that gives the optimum coherence
$ \frac{1}{\sqrt{2^{m-1}}} = \sqrt{\frac{2}{M}}$ for $\Phibu_{k_1, k_2}$
with the highest probability.
However, if one randomly chooses a permutation pair for even $m$,
the most probable coherence of $\Phibu_{k_1, k_2}$
will be $\frac{1}{\sqrt{2^{m-2}}} = \frac{2}{\sqrt{M}}$, which is sub-optimum,
or twice of the optimum coherence in Corollary~\ref{co:min_coh}.

Exploiting the experimental result of $p_r$, 
we can determine the coherence of the spreading matrix $\Phibu$
stochastically.

\begin{thr}\label{th:coh_pr}
In Definition~\ref{def:spread}, assume that
one chooses a permutation set
$\Gamma = \{ \pi_1, \cdots, \pi_L \}$ randomly
for the spreading matrix $\Phibu$. 
Let $\Bbu_{k_1, k_2}$ be the symplectic matrix corresponding to
a permutation pair $\pi_{k_1}$ and $\pi_{k_2}$ in the set $\Gamma$, where $k_1 \neq k_2$,
and $p_r$ be the probability of 
${\rm rank}_2 \left(\Bbu_{k_1, k_2} \right) = r$,
where $r$ is an even integer for $0 \leq r \leq \lfloor \frac{m}{2} \rfloor$.
Then, $\mu \left( \Phibu \right) = \frac{1}{\sqrt{2^r}}$ with probability
\begin{equation*}\label{eq:P_r}
P_r = \left(\sum_{h=r/2} ^{\lfloor m/2 \rfloor} p_{2h}  \right) ^{\frac{L(L-1)}{2}}
- \left(\sum_{h=r/2+1} ^{\lfloor m/2 \rfloor} p_{2h}  \right) ^{\frac{L(L-1)}{2}}.
\end{equation*}
\end{thr}

\noindent \textit{Proof}: 
In order to achieve $\mu \left( \Phibu \right) = \frac{1}{\sqrt{2^r}}$, 
Theorem~\ref{th:rank_coh} requires 
${\rm rank}_2 \left( \Bbu_{k_1, k_2} \right) \ge r$
for any permutation pair $\pi_{k_1}$ and $ \pi_{k_2}$ in $\Gamma$. 
Let $r_1,  \cdots, r_{\tau}$ be the ranks of $ \Bbu_{k_1, k_2} $ 
corresponding to all possible choices of
$(\pi_{k_1}, \pi_{k_2})$ from $\Gamma$,
where $\tau = \binom{L}{2} = \frac{L(L-1)}{2}$.
Then, 
\begin{equation}\label{eq:pri}
{\rm Pr}\left[r_i \ge r \right] = \sum_{h=r/2} ^{\lfloor m/2 \rfloor} p_{2h}, \quad i = 1, \cdots, \tau.
\end{equation}
Under a mild assumption that \eqref{eq:pri} is 
independent for each $i$,
we have 
\begin{equation}\label{eq:prL}
{\rm Pr} \left[ r_1 \ge r, \cdots r_{\tau} \ge r \right] = \prod_{i=1} ^{\tau} {\rm Pr} \left[r_i \ge r \right].
\end{equation}
Finally,
the probability of $r_{\min} = r$ at Theorem~\ref{th:rank_coh} 
is 
\begin{equation*}\label{eq:pr2}
\begin{split}
{\rm Pr}\left[r_{\min} = r \right] & = {\rm Pr} \left [ r_1 \ge r,  \cdots r_{\tau} \ge r \right]  
 - {\rm Pr} \left[ r_1 \ge r+2, \cdots r_{\tau} \ge r+2 \right],
\end{split}
\end{equation*}
which yields $P_r$ 
from \eqref{eq:pri} and \eqref{eq:prL}.
\qed


\begin{cor}\label{co:coh_pr}
If $m$ is odd, a random trial gives
$\Phibu$ of $\mu_{\rm opt} ^{(\rm odd)} \left( \Phibu \right) = \sqrt{\frac{2}{M}}$ 
with probability
\begin{equation*}\label{eq:pr_odd}
P_{m-1} = \left(p_{m-1} \right) ^{\frac{L(L-1)}{2}}.
\end{equation*}
Meanwhile, if $m$ is even, 
the probability that
we obtain $\Phibu$ with $\mu_{\rm opt} ^{(\rm even)} \left( \Phibu \right) = \sqrt{\frac{1}{M}}$
through a random selection is 
\begin{equation*}\label{eq:pr_even}
P_{m} = \left(p_{m} \right) ^{\frac{L(L-1)}{2}}.
\end{equation*}
\end{cor}

\begin{figure*}
	\centering
	\includegraphics[width=1\textwidth, angle=0]{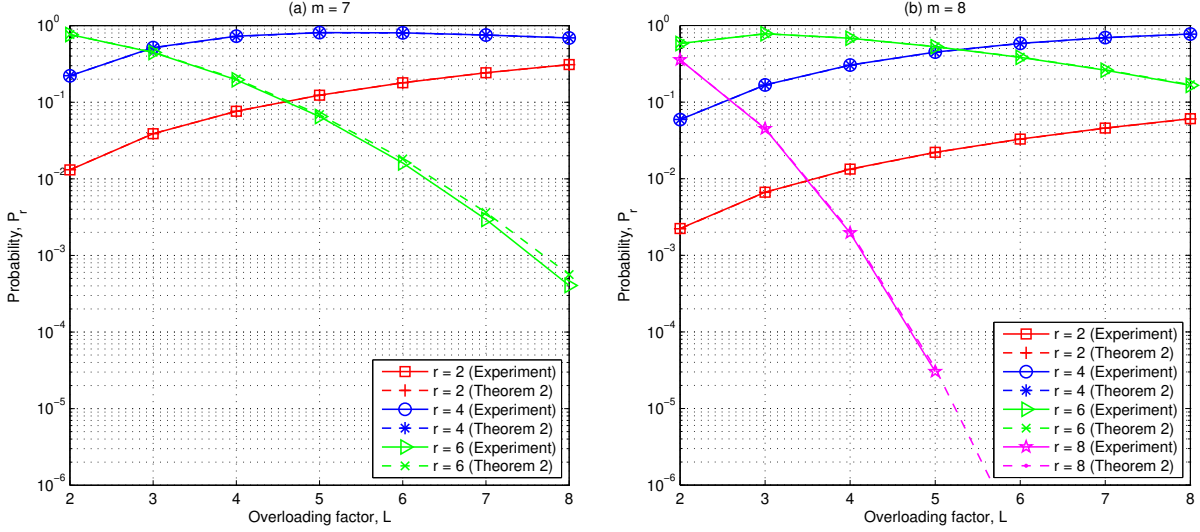}
	\caption{Probabilities of $\mu \left(\Phibu\right) = \frac{1}{\sqrt{2^r}}$ for $M = 2^m$ with $m=7$ and $8$,
		when $L$ permutations are randomly chosen for $\Phibu$.}
	\label{fig:coh_phi}
\end{figure*}

Figure~\ref{fig:coh_phi} displays the probabilities of $\mu \left(\Phibu \right) = \frac{1}{\sqrt{2^r}}$
for various $r$ with respect to the overloading factor $L$, $2 \le L \le 8$, where
$M=2^m$ with $m=7$ and $8$.
In the figure, we computed the theoretical $P_r$ of Theorem~\ref{th:coh_pr} using the experimental $p_r$ of Table~\ref{tb:coh_pair},
whereas the experimental $P_r$ was obtained through $10^7$ random selections of $\Gamma$. 
Figure~\ref{fig:coh_phi} demonstrates that the result of Theorem~\ref{th:coh_pr} well predicts the experimental counterpart,
based on $p_r$ of Table~\ref{tb:coh_pair}. 
It also reveals that if $m = 8$,
the probability $P_m$ of $\mu_{\rm opt} ^{(\rm even)} \left( \Phibu \right) = \sqrt{\frac{1}{M}}$   
becomes extremely low when the overloading factor $L$ is large.
Through further experiments,
we made a similar observation  
that if $L$ is large for even $m$,
a random search is highly unlikely to find a permutation set 
resulting in the optimum coherence.

Next, we investigate the minimum number of random trials 
to find the spreading matrix $\Phibu$ with the optimum coherence.
\begin{thr}\label{th:LT}
When the overloading factor $L$ is given, 
assume that we search for a permutation set $\Gamma$ 
through random trials.
Then, for a small $\epsilon > 0$, the number of trials $T_r$ should be
\begin{equation}\label{eq:T_r}
T_r \ge \frac{\log \epsilon}{\log \left(1-P_r  \right)}
\end{equation}
to achieve $\Phibu$ of $\mu \left( \Phibu \right) = \frac{1}{\sqrt{2^r}}$ 
with probability exceeding $1- \epsilon$,
where $P_r$ is from Theorem~\ref{th:coh_pr}. 
In particular, 
if the number of random trials is 
\begin{equation}\label{eq:T_opt}
T_{\rm opt} \ge \frac{\log \epsilon}{\log \left(1-p_{\rm opt} ^{\frac{L(L-1)}{2} } \right)},
\end{equation}
then $\Phibu$ of $\mu_{\rm opt} ^{(\rm odd)} \left( \Phibu \right) $  
(or $\mu_{\rm opt} ^{(\rm even)} \left( \Phibu \right) $) can be found
with probability exceeding $1- \epsilon$ for a small $\epsilon >0$,
where $p_{\rm opt}= p_{m-1}$ for odd $m$, and $p_{\rm opt} = p_{m}$ for even $m$,
respectively. 
\end{thr}

\noindent \textit{Proof}: 
Theorem~\ref{th:coh_pr} implies that if $\Gamma$ is chosen by a single random trial,
we can achieve $\mu \left( \Phibu \right) = \frac{1}{\sqrt{2^r}}$ with probability $P_r$.
Assuming $T_r$ independent trials, 
we can consider the binomial distribution $\mB (T_r, P_r)$,
where $P_r$ is the probability of $\mu \left( \Phibu \right) = \frac{1}{\sqrt{2^r}}$ at each trial.
Then, 
one can find a permutation set $\Gamma$ 
that yields $\Phibu$ of $\mu \left( \Phibu \right) = \frac{1}{\sqrt{2^r}}$ at least once
with probability
\[
P_1 = 1 - \left(1 - P_r \right)^{T_r}. 
\]
Thus,
\eqref{eq:T_r} is straightforward from 
$P_1 \ge 1-\epsilon$.
Also, \eqref{eq:T_opt} is immediate from \eqref{eq:T_r} and Corollary~\ref{co:coh_pr}.
\qed

\begin{figure}
	\centering
	\includegraphics[width=0.65\textwidth, angle=0]{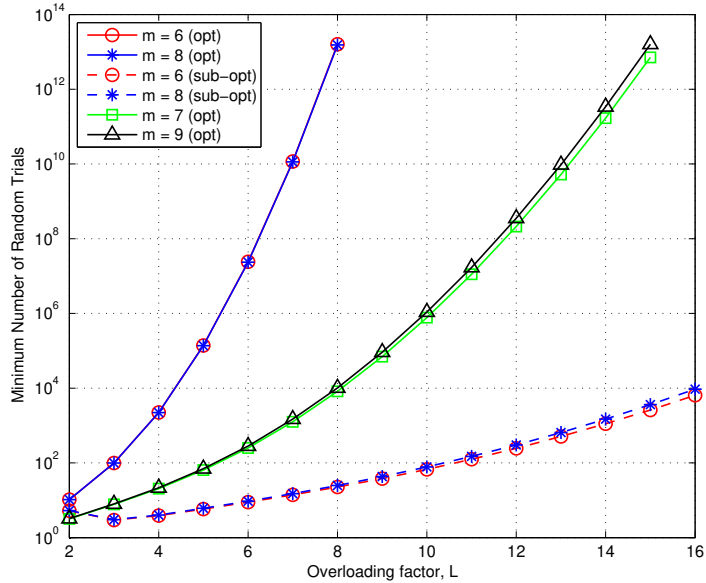}
	\caption{Minimum number of random trials to obtain $\Phibu$ of optimum and sub-optimum coherence with $99 \%$ probability.}
	\label{fig:trials}
\end{figure}

Figure~\ref{fig:trials} displays the minimum number of random trials 
of Theorem~\ref{th:LT} to find $\Phibu$ with optimum and sub-optimum coherence, where $\epsilon = 0.01$.
In the figure, `opt' means the number of random trials for the optimum coherence, 
while `sub-opt' is the number for
the sub-optimum coherence $\mu \left( \Phibu \right) = \frac{2}{\sqrt{M}}$ for even $m$.
The figure shows that if $L \le 10$ for $m=7$ and $9$, 
about $10^6$ random trials can find a permutation set $\Gamma$
with $99\%$ probability that gives the optimum coherence of $\Phibu$.
However, if $m=6$ and $8$, 
the number of random trials increases dramatically as the overloading factor $L$ gets larger.
For example, if $L\ge 8$, 
the number of random trials to find
$\Phibu$ with the optimum coherence
is tremendously large.
Instead, 
it seems relatively easy to obtain
$\Phibu$ with the sub-optimum coherence for $m=6$ and $8$,
since about $10^4$ random trials can find such $\Phibu$ 
with $99\%$ probability
even for $L = 16$.

\begin{table*}[!t]
	\caption{Permutation sets for $\Phibu$}
	\label{tb:pm_set}
	\fontsize{8}{10pt}\selectfont
	\centering
	\begin{tabular}{c|c|c|c|l}
		\hline
		$m$ & $M$ & Overloading factor & Coherence  & Permutation set $\Gamma$   \\
		\hline
		$5$  &  $32$  & $2 \le L \le 8$  & $0.25$  & $(5,4,3,2,1),(3,4,2,5,1),(4,2,5,3,1),(4,3,5,1,2),(4,5,1,3,2),(5,3,1,4,2),$ \\ 
		 	 &     	  &					 &		   & $(5,4,2,1,3),(4,1,2,5,3). $     \\
		\hline
		$6$  &  $64$  & $2 \le L \le 5$  & $0.125$  & $(3,4,5,2,6,1),(6,3,2,4,1,5),(4,1,6,5,2,3),(6,5,3,1,2,4),(5,3,2,1,6,4). $     \\
		     &        & $6 \leq L \le 8$ & $0.25$   & $(3,4,5,2,6,1),(6,4,2,1,5,3),(6,1,4,3,2,5),(4,1,5,6,2,3),(4,2,1,5,6,3),$   \\
		     &        &  				 & 			& $(6,5,3,1,4,2),(6,1,5,3,4,2),(6,2,3,1,5,4). $     \\	
		\hline
		$7$  &  $128$ & $2 \leq L \le 8$ & $0.125$  & $(4,5,1,3,6,7,2),(4,2,5,1,6,7,3),(6,7,1,2,3,5,4), (5,3,6,4,1,7,2),$     \\
        	 &        &  				 & 			& $(6,4,7,3,1,5,2),(4,3,6,7,5,2,1),(6,1,3,2,7,4,5),(6,7,5,1,4,3,2). $     \\	
		\hline
		$8$  &  $256$ & $2 \le L \le 5$  & $0.0625$  & $(4,5,6,1,3,7,8,2),(7,6,8,2,3,1,4,5),(7,1,8,6,4,3,5,2),(6,7,2,3,8,4,1,5), $     \\
    		 &        &  				 & 		     & $(8,3,1,5,2,7,4,6). $     \\	
			&        & $6 \leq L \le 8$ & $0.125$   & $(5,7,4,3,2,8,6,1),(5,7,8,4,6,2,1,3),(5,6,2,7,8,3,4,1),(5,3,1,6,8,7,2,4),$   \\
			&        &  				 & 			& $(8,3,1,7,6,2,4,5),(6,1,3,7,2,8,4,5),(5,1,8,6,7,2,3,4),(8,1,4,6,7,5,2,3). $     \\	
		\hline
		$9$ &  $512$ & $2 \leq L \le 8$ & $0.0625$  & $(8,3,7,4,9,2,5,1,6),(8,4,3,7,2,6,1,9,5),(9,5,4,1,6,8,3,7,2),$     \\
			&        &  				 & 			& $(6,5,8,7,9,3,4,2,1),(4,1,7,6,8,9,2,5,3),(4,8,2,6,9,7,5,3,1), $     \\	
			&        &  				 & 			& $(5,3,7,8,2,1,6,9,4),(5,6,9,3,7,1,8,2,4). $     \\	
		\hline
		$10$  &  $1024$ & $2 \le L \le 5$  & $0.03125$  & $(9,1,6,3,2,8,5,4,10,7),(5,1,9,8,2,10,6,3,7,4),(6,3,8,10,9,7,1,5,4,2),$     \\
	 		  &         &  				   & 		    & $(7,6,8,1,3,2,10,9,4,5),(9,5,3,2,4,8,6,10,7,1). $     \\	
			  &        & $6 \leq L \le 8$ & $0.0625$   & $(5,4,8,1,7,9,10,6,2,3),(8,9,3,4,10,1,6,2,5,7),(6,2,7,8,5,4,3,9,10,1),$   \\
			  &        &  				 & 		       & $(9,10,8,3,4,1,7,2,6,5),(5,8,4,7,9,10,3,6,2,1),(6,4,8,2,7,10,5,9,1,3), $     \\	
			  &        &  				 & 		       & $(3,6,10,4,1,8,9,5,7,2),(8,5,7,2,10,1,6,9,3,4).$ \\
		\hline
	\end{tabular}
\end{table*}

\begin{rem}\label{rm:coh_prob}
	Figure~\ref{fig:trials} shows that if the overloading factor $L$ is large for even $m$, 
	it is less likely 
	to find a permutation set 
	that provides $\Phibu$ with the optimum coherence, 
	whereas 
	a permutation set presenting the sub-optimum coherence 
	can be found with high probability.
	Taking into account this guideline, 
	it is reasonable that we 
	use a permutation set that gives for moderate $L$, say $L \leq 8$,
	\begin{equation}\label{eq:coh_paper}
	\mu \left( \Phibu \right) = \left \{ \begin{array}{lll} \sqrt{\frac{2}{M}}, & \mbox{ if } m \mbox{ is odd and } 2 \leq L \leq 8, \\
	\frac{1}{\sqrt{M}}, & \mbox{ if } m \mbox{ is even and } 2 \leq L \leq 5 , \\
	\frac{2}{\sqrt{M}}, & \mbox{ if } m \mbox{ is even and } 6 \le L \le 8 , \end{array} \right.
	\end{equation}
	where $5 \leq m \leq 10$. 
	Table~\ref{tb:pm_set} presents such permutation sets that give \eqref{eq:coh_paper}, found by $10^6$ random trials,
	where $2 \leq L \leq 8$.
	For odd $m$, if one chooses an arbitrary set of $L$ permutations out of $8$ ones in the table,
	the optimum coherence of $\Phibu$ is guaranteed. 
	For even $m$, if $ L \le 5$, an arbitrary $L$ selection out of $5$ ones provides the optimum coherence for $\Phibu$,
	but if $ 6 \leq L \le 8$, the sub-optimum coherence can be obtained by
	an arbitrary $L$ selection out of $8$ ones in the table. 
	When $L \ge 6$ for even $m$, it is clear that
	the permutations do not correspond to the quadratic forms of a Kerdock code. 
\end{rem}

\section{Simulation Results}

In this section, we examine the performance of binary Golay spreading sequences
for CS-based joint CE and MUD in uplink grant-free NOMA. 
For comparison, we test with random bipolar and Gaussian sequences of length $M=2^m$. 
The random bipolar sequences take the elements of $\pm \frac{1}{\sqrt{M}}$ uniformly at random, 
while each element of the random Gaussian sequences follows the normal distribution 
with zero mean and variance $\frac{1}{M}$.
Also, we test with Zadoff-Chu (ZC) sequences, 
where the sequence length is set as $M_{\rm ZC}$, a prime number closest to $M$.
The corresponding spreading matrix is then obtained by all cyclic shifts of 
the ZC sequences with $L$ randomly chosen, distinct roots.
In simulations, we set $M = 128$ and $256$
for random bipolar, Gaussian, and binary Golay sequences, while
$M_{\rm ZC} = 127$ and $257$ for ZC sequences.
Finally, we assume that a BS accommodates $N = ML$ overloaded devices with the overloading factor $L$, 
where $2 \leq L \leq 8$.

\subsection{PAPR and Coherence}

\begin{figure}
	\centering
	\includegraphics[width=0.65\textwidth, angle=0]{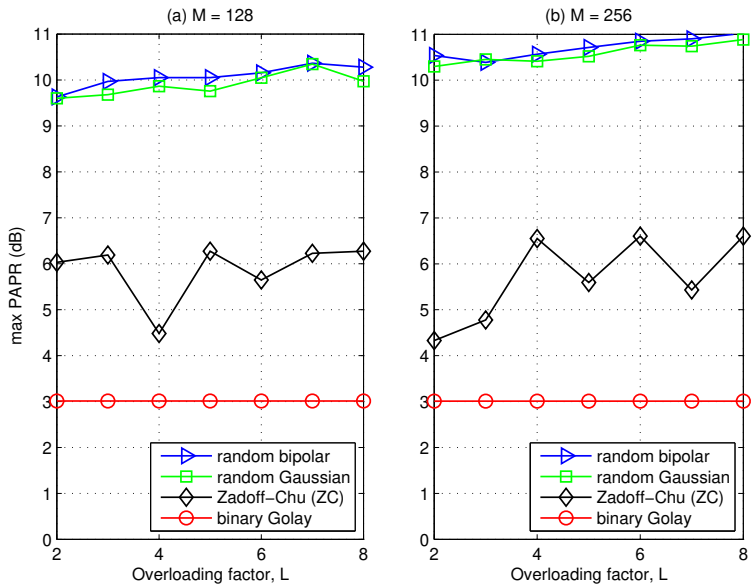}
	\caption{Maximum PAPR of transmitted OFDM signals via various spreading sequences. 
		(a) $M=128$ ($M_{\rm ZC}=127$), (b) $M=256$ ($M_{\rm ZC}=257$).}
	\label{fig:max_papr}
\end{figure}

Figure~\ref{fig:max_papr} shows the maximum PAPR of transmitted OFDM signals,
where QPSK modulated data is spread by various spreading sequences.
In case of random bipolar and Gaussian sequences,
we sketched the smallest values among $100$ random trials for the maximum PAPR. 
As affirmed by Definition~\ref{def:spread}, the PAPR of the binary Golay sequences
is at most $3$ dB, 
which turns out to be significantly lower than those for random bipolar, Gaussian, and ZC sequences.
Thus, the binary Golay spreading sequences allow MTC devices to be equipped with cost-effective power amplifiers
in uplink grant-free NOMA.

\begin{figure}
	\centering
	\includegraphics[width=0.65\textwidth, angle=0]{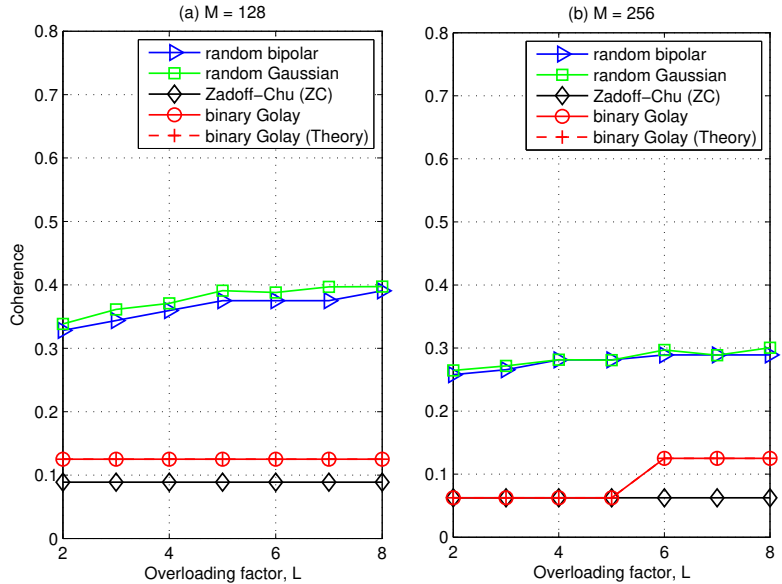}
	\caption{Coherence of spreading matrix $\Phibu$. (a) $M=128$ ($M_{\rm ZC} = 127$), (b) $M=256$ ($M_{\rm ZC}=257$).}
	\label{fig:coh}
\end{figure}

Figure~\ref{fig:coh} displays the coherence of $\Phibu$ for various spreading sequences.
Similar to Figure~\ref{fig:max_papr}, we sketched the smallest coherence among $100$ random trials
for random bipolar and Gaussian spreading sequences. 
Using the permutation sets of Table~\ref{tb:pm_set},
Figure~\ref{fig:coh} confirms 
that the theoretical values of the table
are identical to
the coherence for the binary Golay sequences in the experiment. 
The coherence for ZC sequences is $\frac{1}{\sqrt{M_{\rm ZC}}}$
from the cross-correlation of ZC sequences with distinct roots~\cite{Sarwate:bound}. 
The figure suggests that
the performance of CS-based joint CE and MUD for the binary Golay sequences
might be worse than that for ZC sequences, due to the slightly higher coherence.
But, simulation results in next subsection show that
the performance gap is negligible
if the overloading factor is moderate, e.g., $L \le 6$.

\subsection{Performance of CS-based Joint CE and MUD}

In uplink grant-free NOMA, 
we assume that device activity is uniformly distributed over all $N $ devices.
The data symbol of an active device is QPSK modulated at each time slot.
A frame consists of $J=7$ continuous time slots
for which the device activity remains unchanged. 
To exploit the frame-wise joint sparsity in MMV recovery, 
we use the simultaneous orthogonal matching pursuit
(SOMP)~\cite{Tropp:somp}
for CS-based joint CE and MUD.
Note that the true or average number of active devices in a frame is unknown to BS,
which requires the SOMP to be \emph{sparsity-blind}\footnote{In simulations,
we empirically stop the iteration of this sparsity-blind SOMP
if the maximum of the residual norm is less than $\sqrt{3 \sigma_n ^2 J}$.}
with no prior knowledge of the number.
In simulations, the signal-to-noise ratio (SNR) per device is set as 
$\frac{1}{K} \cdot \frac{ \sum_{t=1} ^J || \ybu ^{(t)} ||_2 ^2}{JM \sigma_n ^2}$,
where $K$ is the true number of active devices in a frame.

\begin{figure}
	\centering
	\includegraphics[width=0.65\textwidth, angle=0]{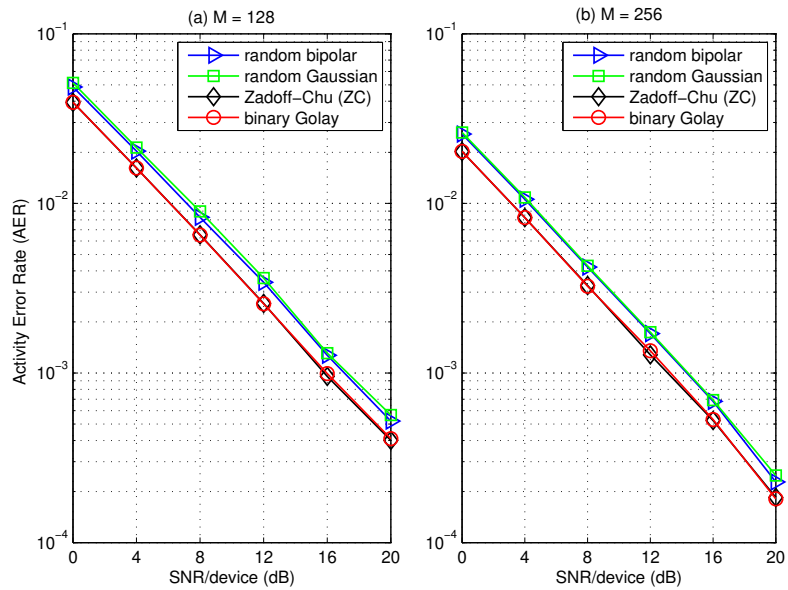}
	\caption{Activity error rates (AER) of CS-based joint CE and MUD over ${\rm SNR}$ per device,
		where $L=4$, $J=7$, and $p_a = 0.1$.
		(a) $M = 128$ ($M_{\rm ZC}=127$),
		(b) $M = 256$ ($M_{\rm ZC}=257$).}
	\label{fig:aer}
\end{figure}

\begin{figure}
	\centering
	\includegraphics[width=0.65\textwidth, angle=0]{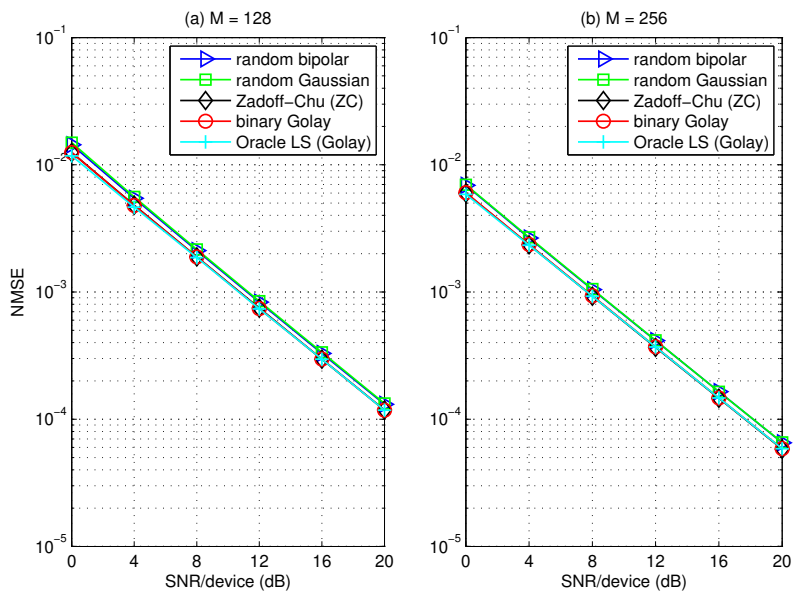}
	\caption{Normalized mean squared errors (NMSE) of CS-based joint CE and MUD over ${\rm SNR}$ per device,
		where $L=4$, $J=7$, and $p_a = 0.1$.
		(a) $M = 128$ ($M_{\rm ZC}=127$),
		(b) $M = 256$ ($M_{\rm ZC}=257$).}
	\label{fig:nmse}
\end{figure}

\begin{figure}
	\centering
	\includegraphics[width=0.65\textwidth, angle=0]{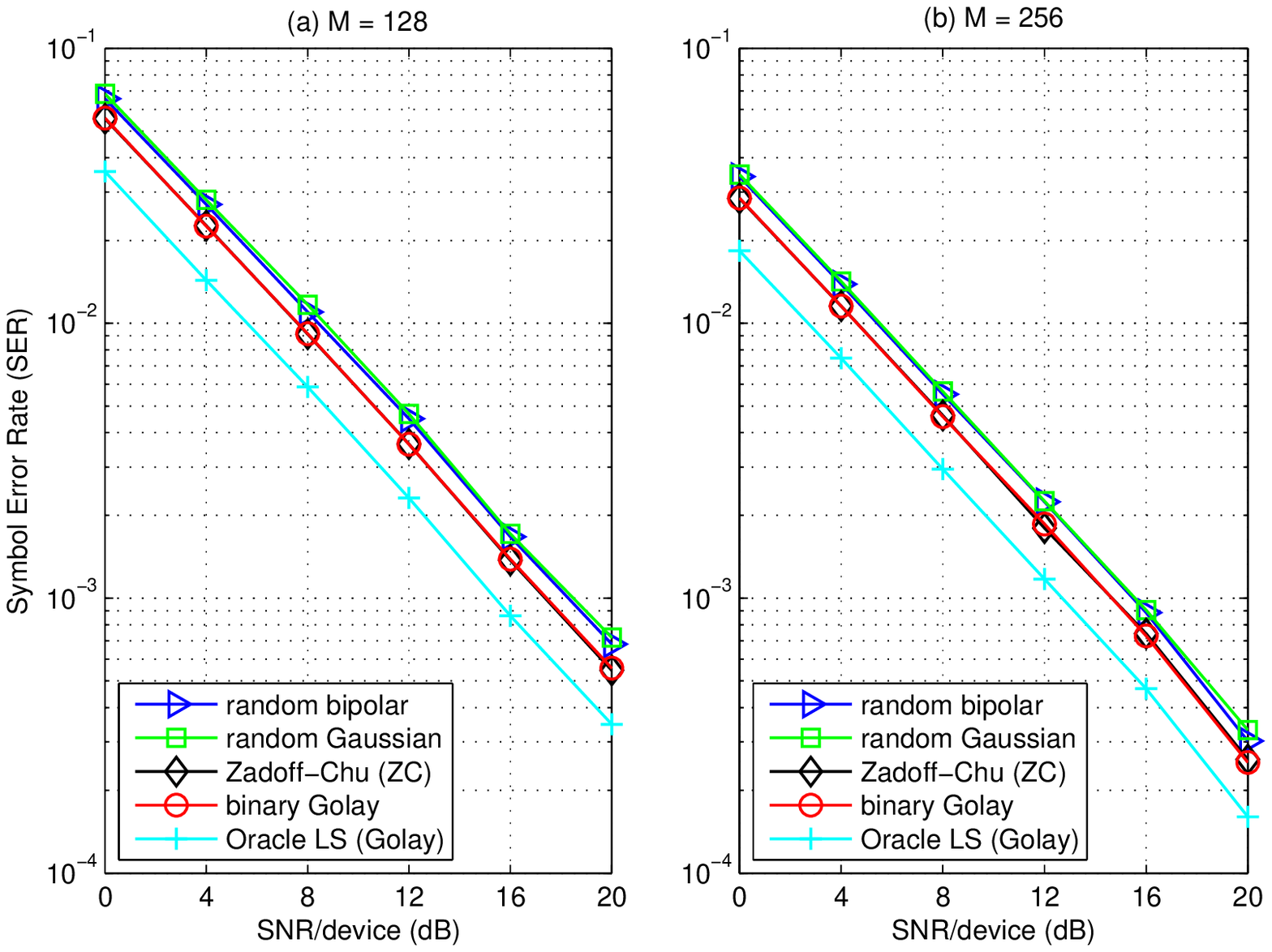}
	\caption{Symbol error rates (SER) of CS-based joint CE and MUD over ${\rm SNR}$ per device,
		where $L=4$, $J=7$, and $p_a = 0.1$.
		(a) $M = 128$ ($M_{\rm ZC}=127$),
		(b) $M = 256$ ($M_{\rm ZC}=257$).}
	\label{fig:ser}
\end{figure}

To evaluate the performance of 
CS-based joint CE and MUD, 
we examine activity error rates (AER) for activity detection, 
normalized mean squared errors (NMSE) for channel estimation, 
and symbol error rates (SER) for data detection, respectively,
over $10^4$ frames.
Figures~\ref{fig:aer}$-$\ref{fig:ser} 
depict the AER, NMSE, and SER over SNR per device, where 
the activity rate is $p_a = 0.1$ and 
the overloading factor is $L=4$.
For example, if $M = 256$ and $L = 4$ with $p_a = 0.1$, 
there are total 
$N=ML=1024$ devices and $\left \lfloor p_a N \right \rfloor = 102 $ ones among them are active on average in a frame. 
In activity detection of Figure~\ref{fig:aer}, 
both undetected and false-alarmed devices are treated as errors.
In Figure~\ref{fig:nmse}, the channel estimation errors are
measured by the normalized mean squared errors (NMSE),
i.e., the average of
$\frac{|| \hbu^\mS - \widehat{\hbu}^{\mS} ||_2 ^2}{|| \hbu^\mS ||_2 ^2}$,
where $\hbu^\mS$ and $ \widehat{\hbu}^{\mS}$ are true and estimated channel vectors,
respectively, for truly active devices.
In Figure~\ref{fig:ser}, 
symbol errors are counted if either
a symbol error occurs from a detected device or the device activity fails to be detected.
As a benchmark, we sketch the NMSE and SER of oracle least squares (LS)
for the binary Golay sequences,
where the true support set $\mS$ is known a priori. 
Note that the benchmark can be considered as the best achievable performance 
for the binary Golay sequences, regardless of CS recovery algorithms.
Figures~\ref{fig:aer}$-$\ref{fig:ser} demonstrate that the
binary Golay sequences outperform random bipolar and Gaussian sequences in AER, NMSE, and SER,
and exhibit the performance similar to those of ZC sequences.

\begin{figure}
	\centering
	\includegraphics[width=0.65\textwidth, angle=0]{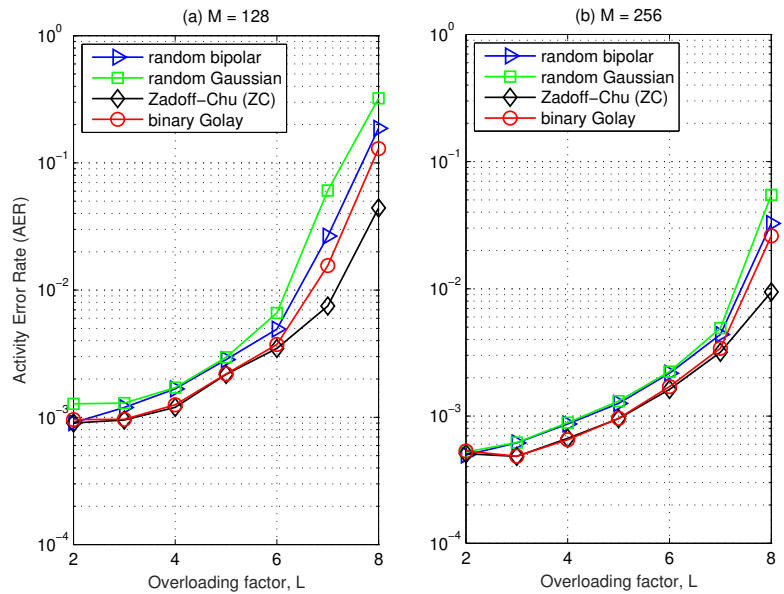}
	\caption{Activity error rates (AER) of CS-based joint CE and MUD over the overloading factor $L$,
		where ${\rm SNR}=15 $ dB, $J=7$, and $p_a = 0.1$.
		(a) $M = 128$ ($M_{\rm ZC}=127$),
		(b) $M = 256$ ($M_{\rm ZC}=257$).}
	\label{fig:aer_L}
\end{figure}

\begin{figure}
	\centering
	\includegraphics[width=0.65\textwidth, angle=0]{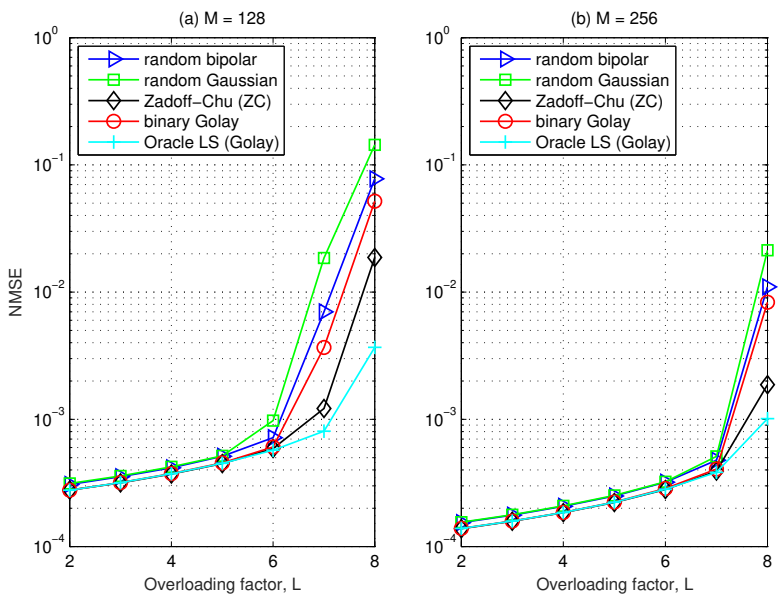}
	\caption{Normalized mean-squared errors (NMSE) of CS-based joint CE and MUD over the overloading factor $L$,
		where ${\rm SNR}=15 $ dB, $J=7$, and $p_a = 0.1$.
		(a) $M = 128$ ($M_{\rm ZC}=127$),
		(b) $M = 256$ ($M_{\rm ZC}=257$).}		
	\label{fig:nmse_L}
\end{figure}

\begin{figure}
	\centering
	\includegraphics[width=0.65\textwidth, angle=0]{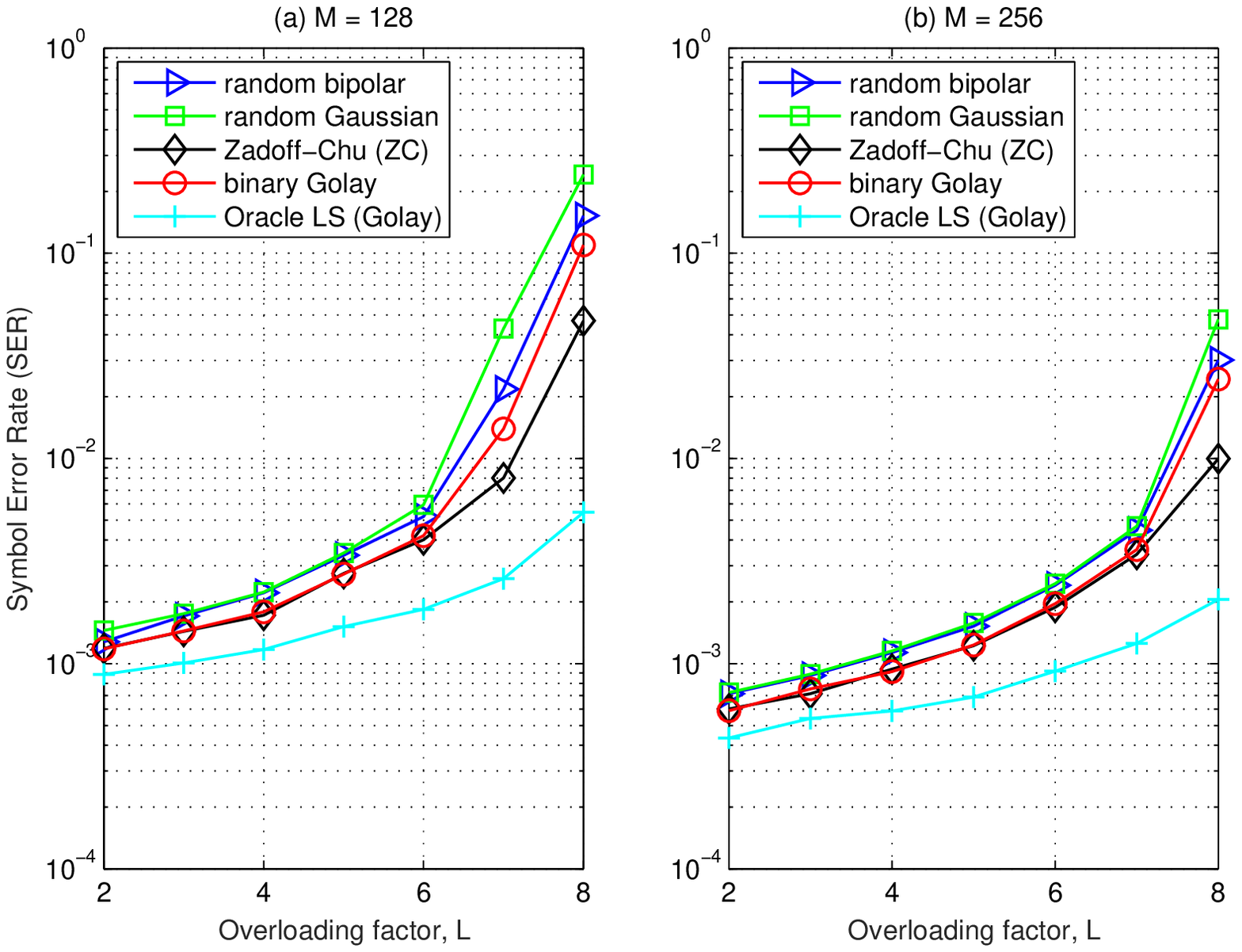}
	\caption{Symbol error rates (SER) of CS-based joint CE and MUD over the overloading factor $L$,
		where ${\rm SNR}=15 $ dB, $J=7$, and $p_a = 0.1$.
		(a) $M = 128$ ($M_{\rm ZC}=127$),
		(b) $M = 256$ ($M_{\rm ZC}=257$).}
	\label{fig:ser_L}
\end{figure}

Figures~\ref{fig:aer_L}$-$\ref{fig:ser_L} sketch
the AER, NMSE, and SER over the overloading factor $L$,
where $p_a = 0.1$ and ${\rm SNR} = 15$ dB per device.
The figures demonstrate that the performance of
the binary Golay spreading sequences is superior to those of random bipolar and Gaussian
sequences.
Compared to ZC sequences, the binary Golay sequences exhibit the similar performance if the
overloading factor is low or moderate, e.g., $L \leq 6$.
However, if $L>6$, we observe that 
the binary Golay sequences 
have a slight performance degradation compared to ZC sequences,
which is due to the higher coherence for large $L$, as shown in Figure~\ref{fig:coh}.



\begin{figure}
	\centering
	\includegraphics[width=0.65\textwidth, angle=0]{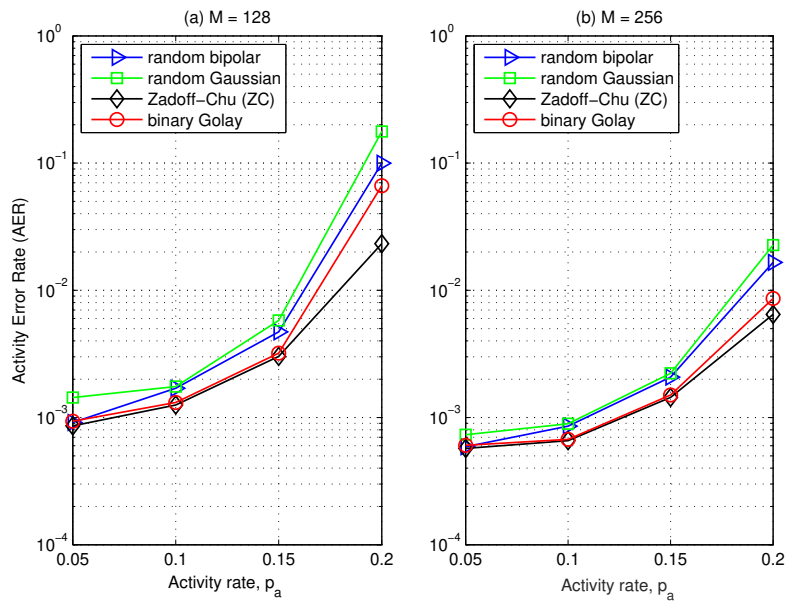}
	\caption{Activity error rates (AER) of CS-based joint CE and MUD over the activity rate $p_a$,
		where ${\rm SNR}=15 $ dB, $J=7$, and $L = 4$.
		(a) $M = 128$ ($M_{\rm ZC}=127$),
		(b) $M = 256$ ($M_{\rm ZC}=257$).}
	\label{fig:aer_pa}
\end{figure}

\begin{figure}
	\centering
	\includegraphics[width=0.65\textwidth, angle=0]{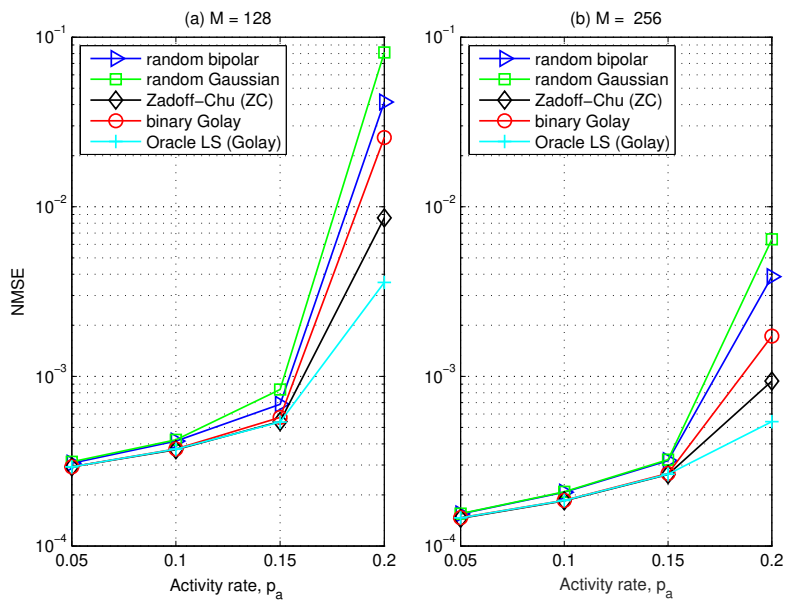}
	\caption{Normalized mean-squared errors (NMSE) of CS-based joint CE and MUD over the activity rate $p_a$,
		where ${\rm SNR}=15 $ dB, $J=7$, and $L=4$.
		(a) $M = 128$ ($M_{\rm ZC}=127$),
		(b) $M = 256$ ($M_{\rm ZC}=257$).}		
	\label{fig:nmse_pa}
\end{figure}

\begin{figure}
	\centering
	\includegraphics[width=0.65\textwidth, angle=0]{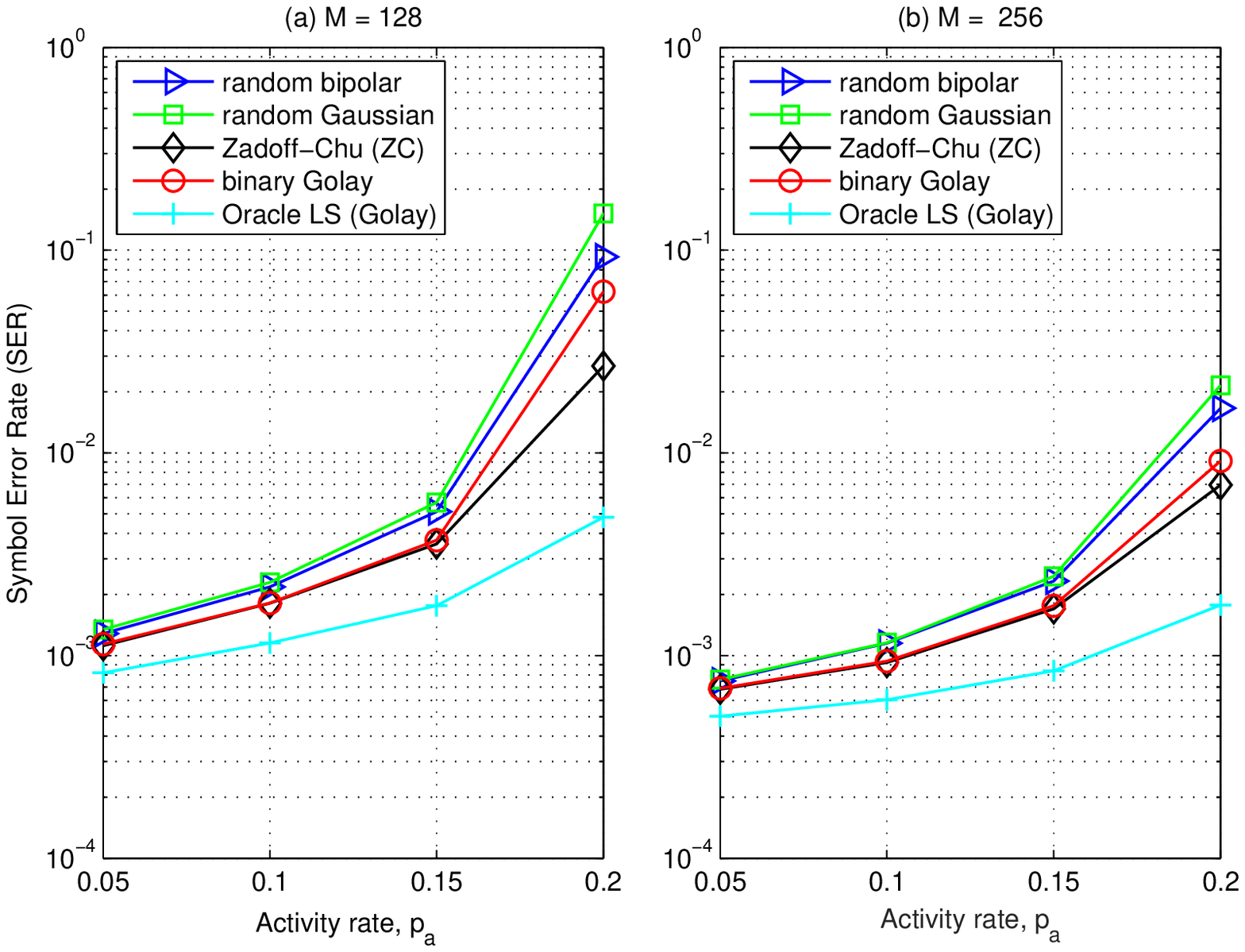}
	\caption{Symbol error rates (SER) of CS-based joint CE and MUD over the activity rate $p_a$,
		where ${\rm SNR}=15 $ dB, $J=7$, and $L=4$.
		(a) $M = 128$ ($M_{\rm ZC}=127$),
		(b) $M = 256$ ($M_{\rm ZC}=257$).}
	\label{fig:ser_pa}
\end{figure}

Figures~\ref{fig:aer_pa}$-$\ref{fig:ser_pa} depict
the AER, NMSE, and SER over the activity rate $p_a$,
where $L=4$ and ${\rm SNR} = 15$ dB per device.
Similar to Figures~\ref{fig:aer_L}$-$\ref{fig:ser_L}, 
the figures demonstrate that the binary Golay sequences outperform random bipolar and Gaussian
sequences.
Compared to ZC sequences,
they show the similar performance for $p_a \leq 0.15$, but
if the activity rate is high ($p_a = 0.2$),
the binary Golay sequences have a performance degradation at $M = 128$, 
which however seems to be mitigated at $M=256$.
It is known that the activity rate of mobile traffic 
does not exceed $ 10 \%$
even in busy hours~\cite{Hong:ra}.
Therefore, Figures~\ref{fig:aer_pa}$-$\ref{fig:ser_pa}
demonstrate that the binary Golay spreading sequences are sufficiently
effective for CS-based joint CE and MUD in uplink grant-free NOMA.

\section{Conclusion}

In this paper, we have studied a set of binary Golay spreading sequences 
to be uniquely assigned to overloaded MTC devices for uplink grant-free NOMA.
Based on Golay complementary sequences, 
each spreading sequence of length $M=2^m$ provides the PAPR of at most $3$ dB for multicarrier transmission.
From the perspective of coding theory, 
each sequence is a second-order coset of the first-order 
Reed-Muller (RM) codes,
where the coset representative is determined by a permutation pattern.
Exploiting the theoretical connection, 
we randomly searched for a set of permutations
that provides the corresponding spreading matrix with theoretically bounded low coherence.
We conducted a probabilistic analysis to find
the minimum number of random trials to obtain the spreading matrix
with optimum or sub-optimum coherence.
Through this analysis, the coherence of the spreading matrix
turned out to be $\mO \left(\sqrt{\frac{1}{M}} \right)$
if the overloading factor is not too high.

Simulation results demonstrated that transmitted multicarrier signals spread by
the binary Golay sequences have the PAPR of at most $3$ dB,
which turned out to be significantly lower than those for 
random bipolar, random Gaussian, and ZC spreading sequences.
It is expected that the low PAPR 
will provide MTC devices with high power efficiency.
Moreover, thanks to the low coherence, 
the binary Golay sequences also
exhibited reliable performance of CS-based joint
activity detection, channel estimation, and data detection
for uplink grant-free NOMA.
The sequences outperform random bipolar and Gaussian sequences
in terms of AER, NMSE, and SER, respectively.
They also showed the similar performance to ZC sequences
for moderate overloading factors and activity rates.
In practical implementation,
the binary Golay spreading sequences have an advantage over ZC sequences
by keeping only two phases regardless of the sequence length,
which can be suitable for low cost devices in MTC. 

While we successfully presented the binary Golay spreading sequences 
through random trials,
a further research will be fruitful
for finding the permutation set in a constructive way.
In theory and practice,
a constructive solution can make a significant contribution 
to non-orthogonal spreading sequences for uplink grant-free NOMA.
Meanwhile, we may replace a CS-based detector with the technique of deep learning (DL)~\cite{Bengjo:DL} for joint CE and MUD.
Employing the binary Golay spreading sequences,
we can train deep neural networks 
for a DL-based detector,
as in other research works~\cite{AMP:Borg}$-$\cite{Chun:DL}, which is our ongoing research topic.

%



\end{document}